\documentclass[12pt]{article}

\usepackage{amsmath,amsfonts,amssymb}
\usepackage{float}

\usepackage{graphicx}

\newcommand{\mathsym}[1]{{}}
\def\id{\protect{{1 \kern-.28em {\rm l}}}}

\def\be{\begin{eqnarray}}
\def\ee{\end{eqnarray}}
\def\ba{\begin{eqnarray}}
\def\ea{\end{eqnarray}}

\def\nn{\nonumber}
\makeatletter
\renewcommand\section{\@startsection {section}{1}{\z@}%
                                   {-3.5ex \@plus -1ex \@minus -.2ex}%
                                   {2.3ex \@plus.2ex}%
                                   {\normalfont\large\bfseries}}
\renewcommand\subsection{\@startsection{subsection}{2}{\z@}%
                                   {-3.25ex\@plus -1ex \@minus -.2ex}%
                                   {1.5ex \@plus .2ex}%
                                   {\normalfont\normalsize\bfseries}}

\makeatother

\def \foot {\footnote}

\def \ha {{1 \over 2}}
\def \td {\tilde}
\def \ci{\cite}

\def \const {{\rm const}}

\def\S{{\mathcal S} }

 \def \J {{\mathcal  J}}

\def\a{\alpha}

\def \a {\alpha}

\def\g{\gamma}
\def\ov{\over}

\def\J{{\mathcal J}}

\def\l{\lambda}

\def\foot{\footnote}

\def\det{\hbox{det}}
\def \ci {\cite}

\def \fo { {1\ov 4}}

\def \l  {\lambda}

\def \const {{\rm const}}

\def \S {{\rm S}}

\def \td {\tilde}

\def \O {{\mathcal O}}

\def \la {\label}

\def \l {\lambda}
\def\foot{\footnote}

\def \adss {$AdS_5 \times S^5~$ }
\newcommand{\rf}[1]{(\ref{#1})}
\def \ov {\over}

\def\cc{\circ}

\def \ha{{1\ov 2}}

\def \no {\nonumber}

\def \J {\mathcal{J}}

\def \S {{\cal S}}
\def \J {{\cal J}}

 \def \bb {\bar \beta}

\def \la {\label}

\def\foot{\footnote}

\def \sql {{\sqrt \lambda}}

\def \adss {$AdS_5 \times S^5$\ }

\def \ov {\over}

\def \varpi {{\rm w}}
\def \OO {{\cal O}}

\def \c {\gamma}

\def \te {\theta}

\def \S  {{\rm S}}

\def \C {{\cal C}}
\def \bea {\be}
\def \eea {\ee}
\def \c {\gamma}

\def \beq {\be}
\def \eeq {\ee}
\def \beqa {\bea}
\def \eeqa {\eea}

\def \c {{\rm a}}

 \def \J {{\cal J}}
 \def \S {{\cal S}}

\def \os  {\OO({\textstyle{ 1\ov \sqrt{\lambda}}})}

 \def \sql {\sqrt{\lambda}}

\def \cc {{c }} 
\def \OO {{\cal O}}
\def \te {\textstyle}
\def \fl {\sqrt[4]{\lambda}}

\def \fo {{\textstyle{1 \ov4}}}
\def \rx {{\rm x}}
\def \hg {{\hat \g}}

\def \C  {{\rm C}}
\def \hC  {{\rm \hat  C}}
\def \dd  {{\rm d}}
\def \bb {{\rm b}}
\def \dDelta {2}
\def \sql {{\sqrt{\lambda}}}
\def \ed {\end{document}}
 \def \an {{\rm an}} \def \nan {{\rm nan}}

\def \td {\tilde}
 
\newcommand{\mc}{\mathcal }
\newcommand{\wh}{\widehat}
\newcommand{\wt}{\widetilde}
    \newcommand{\cJ}{\mathcal{J}}
	\newcommand{\rl}{\sqrt{\lambda}}
		\newcommand{\mZ}{\mathbb{Z}}
		\newcommand{\tsum}{\rm sum}
\newcommand{\tint}{\rm int}
\newcommand{\treg}{\rm reg}
\newcommand{\tsing}{\rm sing}

    \renewcommand{\(}{\left(}
    \renewcommand{\)}{\right)}
\newcommand{\eq}[1]{(\ref{#1})}

\newcommand{\ofrac}[1]{\frac{1}{#1}}  
  
\begin{document}


\overfullrule=0pt
\parskip=2pt
\parindent=12pt
\headheight=0in \headsep=0in \topmargin=0in \oddsidemargin=0in

\vspace{ -3cm}
\thispagestyle{empty}
\vspace{-1cm}

\rightline{Imperial-TP-AAT-2012-07}
\rightline{ITEP-TH-56/12}

\begin{center}
\vspace{1cm}
{\Large\bf  

Quantum corrections
 to spinning superstrings
  in 
  \vskip 6pt
$AdS_3 \times S^3\times M^4$:
      determining  the  dressing phase 
\vspace{1.2cm}
}

\vspace{.2cm}
 {M. Beccaria$^{a}$, F. Levkovich-Maslyuk$^{b}$,  G. Macorini$^{c}$
 and  A.A. Tseytlin$^{d,}$\footnote{Also at Lebedev  Institute, Moscow. }
}

\vskip 0.6cm

{\em 
$^{a}$ Dipartimento di Matematica e Fisica ``Ennio De Giorgi'', \\ Universita' del Salento \& INFN, 
                     Via Arnesano, 73100 Lecce, Italy\\
\vskip 0.16cm

$^{b}$ Institute for Theoretical and Experimental Physics, \\ B. Cheremushkinskaya ul. 25, 117259 Moscow, Russia \\ \& Physics Department, Moscow State University, 119991, Moscow, Russia

\vskip 0.16cm

$^{c}$  
Niels Bohr International Academy and Discovery Center, Niels Bohr
institute,\\ Blegdamsvej 17 DK-2100 Copenhagen, Denmark

\vskip 0.16cm

$^{d}$ Blackett Laboratory, Imperial College, London SW7 2AZ, U.K.
 }

\vspace{.2cm}

\end{center}

\begin{abstract}
 We     study   the   leading 
  quantum string  correction  to the  dressing   phase  in the  asymptotic Bethe Ansatz   system 
  for    superstring in $AdS_3 \times   S^3 \times  T^4$  supported by RR flux. 
  We find  that the phase  should be different from the   phase appearing  in the $AdS_5 \times   S^5$ case. 
  We  use    the  simplest   example   of a  rigid circular  string  with two equal spins in $S^3$  
   and  also consider the general   approach  based on the algebraic curve  description.
  We also discuss    the case of the  $AdS_3 \times   S^3 \times  S^3 \times S^1 $  
  theory  and  find   the dependence of the  1-loop correction to the 
  effective string tension function  $h(\l)$   (expected to enter  the magnon dispersion relation) 
    on the parameters $\a$  related to the ratio of the two 3-sphere radii.
    This correction vanishes in the $AdS_3 \times   S^3 \times  T^4$   case.


\end{abstract}

\allowdisplaybreaks

\newpage
\setcounter{equation}{0} 
\setcounter{footnote}{0}
\setcounter{section}{0}
\renewcommand{\theequation}{1.\arabic{equation}}
 \setcounter{equation}{0}
\setcounter{equation}{0} \setcounter{footnote}{0}
\setcounter{section}{0}

\def \os {O(\textstyle{ {1 \ov (\sqrt{\lambda})^2}} )}
\def \ost {O(\textstyle{ {1 \ov (\sqrt{\lambda})^3}} )}
\def \cc {{c }} 
\def \OO {{\cal O}}
\def \te {\textstyle}
\def \fl {\sqrt[4]{\l}}

\def \ha {{{\textstyle{1 \ov2}}}}
\def \fo {{\textstyle{1 \ov4}}}
\def \rx {{\rm x}}
\def \hg {{\hat \g}}

\def \C  {{\rm C}}
\def \hC  {{\rm \hat  C}}
\def \dd  {{\rm d}}
\def \bb {{\rm b}}
\def \dDelta {2}
\def \sql {{\sqrt{\l}}}

 \def \an {{\rm an}} \def \nan {{\rm nan}}
 \def \nm {\tilde n_{11}}
 \def \tn {{\tilde n}}  \def \uni {{\rm inv}}
\def \ttn  {{\bar n}_{11}}
\def \ed {\end{document}}

\def \sj {\te { S\ov J}} 
\def \sjj   {\te { \S\ov \J}}  

\def \ca {{\rm a}} \def \cb {{\rm b}} \def \cc {{a}}  \def \ct {{\rm  v }}  
\def \te {\textstyle} \def \cm {{\rm b}}

\def \ccc {{\tilde a}}

\def \adt {$AdS_3 \times S^3 \times T^4$ }
\def \adst {$AdS_3 \times S^3 \times  S^3 \times S^1$ }

\def \v {h} \def \u {\td h} 
\def \mksl {\mathfrak{sl}(2)}
\def \mksu {\mathfrak{su}(2)}
\def \ccm {{\bar b}}
\tableofcontents

\section{Introduction and summary}

Recent remarkable    progress in uncovering   integrable   structure   behind  the spectrum   of  quantum strings   
in \adss  \ci{Beisert:2010jr}  which was   much aided by  duality to $\mathcal N=4$ supersymmetric gauge theory   raises 
the question     about  applying similar   integrability-based  methods  (algebraic curve   description of finite-gap solutions, 
its discretisation and magnon scattering S-matrix  as  guides  towards asymptotic Bethe ansatz (ABA), its TBA   generalisation,  etc.)
also in the similar but less   supersymmetric ``low-dimensional'' cases of 
superstring in    $AdS_3 \times   S^3 \times M^4$ 
and $AdS_2 \times   S^2 \times M^6$   supported by R-R fluxes.
In these cases   the dual   conformal theories are poorly understood and thus one has  less 
data in trying to fix  the structure  of the corresponding Bethe ansatz.

The first important step    was  made in \cite{Babichenko:2009dk} 
   where the set   of   ABA  equations    was proposed  for the spectrum of strings  on  $AdS_3 \times   S^3 \times T^4$  and 
   $AdS_3 \times   S^3 \times S^3\times S^1$  described by  the GS superstring 
    action on  the supercosets 
   $PSU(1,1|2) \times PSU(1,1|2)  / SU(1,1) \times SU(2) $   and  
$D(2,1; \a) \times  D(2,1; \a)  / SU(1,1) \times SU(2)\times SU(2)  $.
The  first  model  may be viewed as a special case  of the second: 
if the   radius of $AdS_3$ is set to 1, then the radii of the two   3-spheres   can be parametrized as\foot{The 
meaning of the relation $1 = R_+^{-2} +  R_-^{-2}$  between   the three radii  can  be easily understood as follows.
The   $AdS_3 \times   S^3 \times S^3\times S^1$   background supported by  the   R-R  3-form background  
is a  type IIB 
supergravity solution related by  S-duality to the  same  metric supported  by the NS-NS  3-form flux  (see, e.g. \ci{adsss}).
The corresponding  string sigma model is   simply  $SL(2,R) \times SU(2) \times SU(2) \times SO(2)$   WZW 
model (with world-sheet supersymmetry if treated in NSR  approach).  The dilaton equation of motion 
(with constant dilaton) is then the total central charge condition relating the three (shifted)  levels, i.e. 
$- { 3\ov k_{sl(2)}} + { 3\ov k_{su(2)_+}} + { 3\ov k_{su(2)_-}} =0$. Since the levels are proportional to the radii, 
the  above relation follows.} 
$R_+^2 = \a^{-1}, \ R^2_- = ( 1 - \a)^{-1}$, i.e.  the   $AdS_3 \times   S^3 \times T^4$  model  with $R_2=\infty$ corresponds to $\a=1$.

The starting point was  the classical    integrable   supercoset  sigma model   and 
the discretisation of the corresponding  finite-gap   equations    following closely  the 
analogy  with  the \adss case \ci{Arutyunov:2004vx}   (see   \cite{Zarembo:2010yz}).
It was  conjectured  in    \cite{Babichenko:2009dk} 
   that the corresponding dressing phase   should be the same  BES phase   \cite{Beisert:2006ez}  as in the \adss  case. 

Further   elaborations   of the proposed   ABA system appeared in
 \ci{David:2010yg,OhlssonSax:2011ms,Sax:2012jv,Ahn:2012hw}. 
 In particular, it  was noted  in  \ci{Ahn:2012hw} that due to different algebraic   structure   here one cannot
 fix the   dressing  scalar   factors  in the magnon S-matrix  using crossing symmetry constraints as was  done 
 \ci{Janik:2006dc,Volin:2009uv,Vieira:2010kb}    in the \adss   case, but until very recently it was
  assumed  that the original conjecture 
 of \cite{Babichenko:2009dk}   that the phase should be given by the BES expression    should be 
    correct.\foot{While the present  paper was in preparation, there appeared a preprint 
 \ci{Borsato:2012ud}   where it   is  claimed that there should    be several scalar phase factors   and  they may differ from the BES  expression.} 
 
 The   aim   of the   present paper is  to   suggest a proposal for  the   leading 
  quantum string  correction   to the  ``classical'' AFS  phase  in the   ABA  system 
   of  \cite{Babichenko:2009dk,OhlssonSax:2011ms} 
  following  the same   first-principles approach as originally   used  in \adss  case  \ci{Beisert:2005cw,Hernandez:2006tk,Freyhult:2006vr,Gromov:2007cd}, 
  i.e.   
  by comparing the ABA predictions to the   quantum string  and algebraic curve computations of 
  the  1-loop  corrections to   semiclassical string   energies.\foot{Some semiclassical computations for superstrings in $AdS_3 \times S^3 \times M^4$ appeared  
   earlier    in \ci{Iwashita:2011ha,Forini:2012bb,Rughoonauth:2012qd,Sundin:2012gc}.}
 We will   study the simplest example   of rigid circular  string  with two equal spins in $S^3$  \cite{Frolov:2003qc}
  (and also closely  related,  via an analytic continuation,  case  of $(S,J)$ folded long string \cite{Frolov:2002av})
  and also consider more general algebraic curve approach. 
  Our conclusion is    that   the 
  phase  in the ABA of \cite{Babichenko:2009dk,OhlssonSax:2011ms}     requires a modification from the standard  \adss    form  of 
   \ci{Beisert:2005cw,Hernandez:2006tk}, i.e. the  ABA   for the 
   $AdS_3 \times   S^3 \times  M^4$ theory can not have the standard    BES phase. 
   
In more details, the  phase for the scattering of two magnons with momenta $p_{j}$ and $p_{k}$ in the \adss theory
can be written as \cite{Beisert:2005cw,Beisert:2005wv}
\be
\label{BES}
\vartheta(p_{j}, p_{k}) = 2\,\mathop{\mathop{\sum_{r=2}^{\infty}}_{s\ge r+1}}_{r+s\ \rm odd}
c_{r,s}(\lambda)\,\Big(\frac{\lambda}{16\,\pi^{2}}\Big)^{\frac{r+s-1}{2}}\,\Big[
q_{r}(p_{j})\,q_{s}(p_{k})-q_{s}(p_{j})\,q_{r}(p_{k})
\Big] \ . 
\ee
Here, 
 $q_{n}(p)$ is the elementary magnon $n$-th charge. The strong coupling expansion of the coefficient  functions
$c_{r,s}(\lambda)$ is
\be
\label{cexp}
c_{r,s}(\lambda) =c_{r,s}^{(0)}
+\frac{1}{\sqrt\lambda}\,c_{r,s}^{(1)}+\dots, \la{cc} 
\ee
where $c_{r,s}^{(0)}=\delta_{r+1,s}$ is the AFS contribution \cite{Arutyunov:2004vx} and the one-loop correction is the
HL  phase  found for $r=2$, $s=3$  in \cite{Beisert:2005cw}  and then  in general in \cite{Hernandez:2006tk}. It  is non-vanishing 
 for odd $r+s$,
\be
\label{crsintro}
c_{r,s}^{(1)} = -8\,\frac{(r-1)\,(s-1)}{(r+s-2)\,(s-r)}, \la{13} 
\ee
and  reproduces  the  ``non-analytic''   part of the 1-loop   correction to  $SU(2)$ circular    string energy  \cite{Beisert:2005cw}
\be
\delta E^{AdS_{5}}_1 = 
\frac{1}{\sqrt{\J^{2}+m^{2}}}\,\Big(
m^{2}+2\,\J^{2}\,\log\frac{\J^{2}}{\J^{2}+m^{2}}-\frac{\J^{2}-m^{2}}{2}\,\log\frac{\J^{2}-m^{2}}{\J^{2}+m^{2}}
\Big). \la{14} 
%
%
\ee
Our analysis of the same $SU(2)$ circular string  with two equal spins  in $R_t \times S^{3}\subset AdS_{3}\times S^{3}\times T^{4}$
suggests  that the corresponding non-analytic term that should be reproduced by the  dressing contribution is instead
\beq
\label{eq:FedorAdS3}
 \delta E^{AdS_{3}}_1=\frac{1}{\sqrt{\cJ^2+m^2}}\Big(m^2+\cJ^2 \log \frac{\cJ^2}{m^2+\cJ^2}\Big).
\eeq
This expression is indeed found when the 
$c_{r,s}^{(1)}$ coefficients for the  LL or RR scattering\foot{L and R stand for  the left and right moving sectors
   \ci{Borsato:2012ud} .}
take the following  new form 
\be
\label{crsbarintro}
c^{(1)}_{r,s} = 2\,\frac{s-r}{r+s-2}\ , \la{new} 
\ee
{\it provided}    also that   the summation in (\ref{BES})  now starts from $r=1$.  
This constitutes our proposal   for the 1-loop dressing phase coefficients. 

Below we show   that the 
 coefficients  \rf{new}    
are consistent     with the  string  prediction for the   $SU(2)$ circular string energy. 
We  arrive at \rf{new}    using 
the  semiclassical algebraic curve approach to the  derivation of the  dressing phase  \ci{Gromov:2007cd}. The   
circular string case serves as    a guide  to  how to resolve the    regularization ambiguity
of  the  algebraic  curve approach.  The expression \rf{new}   follows after requiring 
the antisymmetry of  the  coefficients $c^{(1)}_{r,s} $, which is shown to be consistent    with the string 
result  \rf{eq:FedorAdS3}. 

Although the example of the circular string solution does not 
test the mixed LR or RL scattering, we  also propose   that the scattering phase between the 
opposite chirality magnons  takes also the above general form \rf{BES} 
(again with summation in \rf{BES} starting from $r=1$), but with the coefficients 
\be
\overline c^{(1)}_{r,s} = -2\, \frac{r+s-2}{s-r}. \la{17}
\ee
Additional tests of these expressions for $c^{(1)}_{r,s}$ and $\overline c^{(1)}_{r,s}$   
would certainly be important. 

We also   consider the more general $\alpha$-dependent integrable 
 model based   on $AdS_{3}\times S^{3}\times S^{3}\times S^{1}$.
 Here 
we do not attempt to  fix the dressing phase in ABA in a systematic way 
but   compute  the non-analytic (i.e. dressing-related) 
contribution to the one-loop energy for the corresponding  generalized $SU(2)$ circular string. Remarkably, 
this correction  turns out to be 
{\it independent} of $\alpha$ when  written in terms of the   effective string  tension   $h(\lambda)$
that   has   the following strong coupling expansion  
\be
h(\lambda) = \frac{\sqrt\lambda}{4\,\pi}+ \c +\mathcal O\Big(
\frac{1}{\sqrt \lambda}\Big)\ , \ \ \ \ \ \ \ \ \ \         \c_{_{ \rm AdS_{3}\times S^{3}\times S^{3}\times S^{1} }} = \frac{\alpha\log\alpha+(1-\alpha)\log(1-\alpha)}{4\,\pi} \ .  \la{18}
\ee
This   function  
 is expected to  enter the corresponding  magnon dispersion relation  and thus 
 appear in the Bethe Ansatz. Notice that the 1-loop 
shift in \rf{18}   vanishes at $\alpha=0,1$  when we go back to the $AdS_{3}\times S^{3}\times T^4 $  case. 

This paper is organized as follows. 
We shall start in section 2 with a brief review of the ABA equations of   \cite{Babichenko:2009dk,OhlssonSax:2011ms}.  
  We shall then  consider    the  constraints   on the leading quantum string correction to the   dressing phase   for the \adt  case 
  that follow   from   the  expression for the non-analytic term in the 
  1-loop   quantum correction to the  $SU(2)$ circular    string energy  as an input  (section 3). 
  
Next,  in section 4 we shall  discuss   the  algebraic curve setup   \ci{Gromov:2007cd},     finding 
 a non-antisymmetric  expression for the  coefficients  $c_{r,s}$  in the phase,
 apparently  contradicting the requirement following from the discrete form of the Bethe  Ansatz. 
In section 5  we shall     compare the present computation with the one in the \adss case 
  \ci{Beisert:2005cw,Hernandez:2006tk}  and   point out a 
   mismatch between the  standard string and the  algebraic curve  regularizations.
In section 6 we shall  show   that  requiring the antisymmetry of the  $c_{r,s}$ coefficients 
resolves disagreement  between the algebraic curve   approach and 
   string computation in section 3 and  leads to 
our proposal for the coefficients in  \rf{new}. 

In section 7  we shall derive the  non-analytic (dressing)  part of the 1-loop energy  for the  $SU(2)$ 
 circular string case in the \adst case,  pointing out the  role 
  of the  1-loop shift in \rf{18}   and discuss the  $\a\to 1$   limit. 
 
 In Appendix A  we shall  consider  the  $(S,J)$    folded  string with large spins 
 and determine the corresponding coefficients   $c_{1,s}$    in the phase that 
agree with the ones in the $SU(2)$ case. 

\def \no {\nonumber}

\renewcommand{\theequation}{2.\arabic{equation}}
\setcounter{equation}{0}
\section{Asymptotic Bethe  Ansatz  equations for $AdS_{3} \times  S^{3} \times  T^{4}$ model}

As discussed in \cite{Babichenko:2009dk}, type IIB   GS superstring theory 
on $AdS_{3}\times S^{3}\times S^{3}\times S^{1}$  space  with RR 3-form flux  reduces, in a particular 
$\kappa$-symmetry   gauge, to  a supercoset sigma model  which 
is classically integrable. 
The  string   theory  on the  $AdS_{3} \times S^{3} \times T^{4}$ background  with RR flux can be  formally   treated as the
 limiting case ($\alpha=1$)  of the $AdS_{3}\times S^{3}\times S^{3}\times S^{1}$ supercoset  model. 

From the general classical integrability  structure of   $\mathbb{Z}_{4}$ symmetric (super)cosets (see  \cite{Zarembo:2010yz}), one
 can derive the finite gap equations   which  may be written entirely in terms of the group-theory data. 
 Discretizatization  of these   finite gap equations  leading to the 
  associated quantum Bethe equations was  proposed in \cite{Babichenko:2009dk} for the symmetric point ($\a={1 \ov 2}$) 
where the radii of the two 3-spheres are equal and also for the limiting case ($\alpha=1$)   
of $AdS_{3}\times S^{3}\times T^{4}$. 
It should be   mentioned  that  from the point of view of the integrability structure  the limit $\alpha\to 1$ is 
a non-trivial  one \cite{Sax:2012jv}.

In order to fix the notation,  here we shall  briefly   review  the form of the quantum Bethe equations for the 
 case of $AdS_{3}\times S^{3}\times T^{4}$  that we will be mostly considering here. 
The starting point is the Dynkin diagram of the 
$PSU(1,1|2)\times PSU(1,1|2)/(SU(1,1)\times SU(2))$ supercoset. It contains 3+3 nodes associated to the left/right
moving sectors. The quantum Bethe equations are written in terms of the Bethe roots in the spectral plane $x_{i, \ell}$
where $i=1, 2, 3, \overline{1}, \overline{2}, \overline{3}$, $\ell=1, \dots, K_{i}$ and the parameters $x^{\pm}$
 defined by the Zhukovsky map
\be
x^{\pm}+\frac{1}{x^{\pm}} = x+\frac{1}{x}\pm\frac{i}{2\,h(\lambda)}\ . \la{zh}
\ee
 Here the function $h(\lambda)$  cannot determined by the integrability alone. 
The asymptotic Bethe equations  are given by 
\ba
&&\ \ \ \ \ \ \ \  1 = \prod_{k}\frac{x_{1,j}-x_{2,k}^{+}}{x_{1,j}-x_{2,k}^{-}}
\prod_{k}\frac{1-\frac{1}{x_{1,j}\,x_{\overline 2,k}^{+}}}{1-\frac{1}{x_{1,j}\,x_{\overline 2,k}^{-}}}\ ,\la{mi} \\
\label{middle}
&&  \Big(\frac{x_{2,j}^{+}}{x_{2,j}^{-}}\Big)^{L} =
\prod_{k\neq j}\frac{x_{2,j}^{+}-x_{2,k}^{-}}{x_{2,j}^{-}-x_{2,k}^{+}}
\frac{1-\frac{1}{x_{2,j}^{+}x_{2,k}^{-}}}{1-\frac{1}{x_{2,j}^{-}x_{2,k}^{+}}}\,
\sigma^{2}(x_{2,j}, x_{2,k})\no  \\
&&  \times \prod_{k}\frac{x_{2,j}^{-}-x_{1,k}}{x_{2,j}^{+}-x_{1,k}}
 \prod_{k}\frac{x_{2,j}^{-}-x_{3,k}}{x_{2,j}^{+}-x_{3,k}}
\prod_{k}\frac{1-\frac{1}{x_{2,j}^{-} x_{\overline 1, k}}}{1-\frac{1}{x_{2,j}^{+} x_{\overline 1, k}}}
\prod_{k}\frac{1-\frac{1}{x_{2,j}^{-} x_{\overline 3, k}}}{1-\frac{1}{x_{2,j}^{+} x_{\overline 3, k}}}
\prod_{k}\sigma^{-2}(x_{2,j}, x_{\overline 2, k}) \\
&&\ \ \ \ \ \ \ \  1 = \prod_{k}\frac{x_{3,j}-x_{2,k}^{+}}{x_{3,j}-x_{2,k}^{-}}
\prod_{k}\frac{1-\frac{1}{x_{3,j}\,x_{\overline 2,k}^{+}}}{1-\frac{1}{x_{3,j}\,x_{\overline 2,k}^{-}}}\ , \la{mii}
\ea
together with other three equations with $(1,2,3)\leftrightarrow(\overline 1, \overline 2, \overline 3)$  which are are identical
in stucture to the above  apart from the l.h.s of the middle equation  that is reversed, i.e. is  
$\big(\frac{x_{\overline 2,j}^{-}}{x_{\overline 2,j}^+}\big)^{L}$.

The dressing phase  factor $\sigma^2 = e^{i \vartheta} $ 
has the  AFS \ci{Arutyunov:2004vx}  limit at leading order   in strong coupling  
\be
\sigma_{\rm AFS}(x_{j}, x_{k})=\frac{1-\frac{1}{x_{j}^{-}\,x_{k}^{+}}}{1-\frac{1}{x_{j}^{+}\,x_{k}^{-}}}
\,\Big(
\frac{x_{k}^{-}x_{j}^{-}-1}{x_{k}^{-}x_{j}^{+}-1}\,
\frac{x_{k}^{+}x_{j}^{+}-1}{x_{k}^{+}x_{j}^{-}-1}
\Big)^{i\,h\, (x_{k}+\frac{1}{x_{k}}-x_{j}-\frac{1}{x_{j}})}  
\ee
as required  in order to  match  the classical finite-gap equations. 
Apart from this constraint and unitarity, nothing is known  a priori 
about the dressing phase: it   should be  determined by the dynamics of the integrable system, i.e.  its specific form  is not fixed just 
by  the symmetry structure. 

The expression for the quantum string energy  and the momentum constraint are
\be
E = 2\,h(\l) \,\sum_{i=2, \overline 2}\sum_{\ell=1}^{K_{i}}\Big(\frac{1}{x^{+}_{i,\ell}}-\frac{1}{x^{-}_{i,\ell}}\Big),
\qquad\qquad  \prod_{\ell=1}^{K_{2}}\frac{x^{+}_{2,\ell}}{x^{-}_{2,\ell}}\,\prod_{\ell=1}^{K_{\overline 2}}\frac{x^{-}_{\overline 2,\ell}}{x^{+}_{\overline 2,\ell}}=1.
\ee
These  above equations describe 
bound states of $4_{B}+4_{F}$ massive  magnon modes with mass 1. 
From comparison with  semiclassical string theory  it  follows that the relation between $h$ and $\lambda$ at strong coupling is 
\be
h(\lambda) = \frac{\sqrt\lambda}{4\,\pi}+\mathcal O(1), \qquad\qquad  \lambda\gg 1\ .
\ee
As we shall   find in the next section,   the $\mathcal O(1)$ 1-loop   correction  here  vanishes in the \adt case.

\renewcommand{\theequation}{3.\arabic{equation}}
\setcounter{equation}{0}
\section{One-loop correction to   energy of  circular string with two equal spins in $S^{3}$: fixing  leading quantum term in the dressing phase}

Here  we  shall   present the calculation of the one-loop correction to energy for the 
 rigid circular string with two equal spins $J_1=J_2 $  in $S^3$  part of $AdS_{3}\times S^{3}\times T^{4}$.  
 Similar   computation in the 
 $AdS_{5}\times S^{5}$ case can be found in \cite{Frolov:2003qc,Beisert:2005cw}.
 Following  \cite{Beisert:2005cw}, we shall extract the  so-called {\em non-analytic}   part of the 1-loop correction 
 that should be  arising    from the dressing phase in the ABA
 and thus  fix the subleading strong-coupling  part of the coefficients $c_{r,s}$ in the phase. 

\subsection{Non-analytic  term in one-loop    string energy}

The classical solution we consider here is exactly the same as in the $AdS_5 \times S^5$ case: the motion in the $S^3$ is described by
\beq
	X_1+iX_2=\frac{1}{\sqrt{2}}e^{i\cJ \tau+im\sigma},\ \ \ \ \ \ \ \   X_3+iX_4=\frac{1}{\sqrt{2}}e^{i\cJ \tau-im\sigma} 
\eeq
where $X_k$ are the  embedding coordinates on $S^3$ 
and the  $AdS_3$ part of the solution is 
\beq
	Y_3+iY_0=e^{i\kappa\tau}, \ \ \ Y_1=Y_2=0  \ ,  \ \ \ \ \ \ \ \ \   -Y_0^2+Y_1^2+Y_2^2-Y_3^2=-1 \ . 
\eeq
Here  the spins are $J_1 = J_2 = \ha \sql \J$, \  $m$ is the winding number   and 
 the classical energy of this string is 
\beq
	E_{0}=\sqrt{\lambda}\kappa \ , \ \ \ \ \ \ \ \  \kappa=\sqrt{\cJ^2+m^2} \ . \la{33} 
\eeq
The 1-loop correction to the energy is given by the sum of fluctuation frequencies:
\beq
	E_1=\frac{1}{2\kappa}\sum\limits_{n\in\mZ}  \big(
	\omega_n^{B}- \omega_n^{F }  \big) \ . \la{34}
\eeq
 The individual frequencies in the $AdS_5\times S_5$ case were given 
  in \cite{Frolov:2004bh}. The 1-loop correction in the   $AdS_3\times S^3\times T^4$ 
      case is obtained from the $AdS_5\times S_5$ one by a simple truncation -- we 
      remove two bosonic frequencies that correspond to fluctuations in the transverse directions of $S^5$, and halve the AdS and fermionic 
  contributions.
  There will also be four bosonic and four fermionic massless modes coming from the $T^4$; 
  we will  not write them out explicitly as their contributions cancel each other.
   We are then left with two bosonic frequencies that come from the $S^3$ in which the string is rotating,
\beqa
\label{su2S3omega}
	\omega_n^{B }&=&\Big[n^2+2\kappa^2-2m^2\pm2\sqrt{(\kappa^2-m^2)^2+\kappa^2n^2}\Big]^{1/2},
	\eeqa
two frequencies from the AdS part 
\beq
	\omega_n^{B }=\sqrt{n^2+\kappa^2},
\eeq
and four fermionic frequencies
\beq
	\omega_n^{F }=   \sqrt{n^2+\kappa^2-m^2} \pm \const \ . 
\eeq
The additive constant shifts are irrelevant as they will cancel in the result for  \rf{34}  which is 
\beqa
\label{E1eHL}
&&	\ \ \ \ \ \ \ \ \ E_1=\sum\limits_{n\in\mZ}^{} e(n) \ , \\
\label{eT4}
&&	e(n)
	= \sqrt{1+ \frac{(n+\sqrt{n^2-4 m^2})^2}{ 4({\cal J}^2+m^2)}}
	+\sqrt{1+ \frac{n^2}{ {\cal J}^2+m^2}}
	 - 2 \sqrt{1+ \frac{n^2-m^2 }{ {\cal J}^2+m^2}} \, .
\eeqa
It is straightforward to check that this   sum is UV finite.

In the $AdS_5\times S^5$ case, the 
computation of the large $\cJ$ expansion of the 1-loop energy 
 made it possible to partially fix the coefficients of the leading quantum correction to the AFS dressing phase. 
 To study the large $\cJ$ expansion in the  present 
  case we will use the same  method as  in \cite{Beisert:2005cw}.
   When expanding $e(n)$ at large $\cJ$ one gets terms with 
   divergent sums over $n$;
    separating out the convergent (i.e. regular) and divergent (i.e. singular)
     parts we can write $e(n)=e^{\tsum}_{\treg}(n)+e^{\tsum}_{\tsing}(n)$. 
     To deal with the singular part we define $e^{\tint}(x)=e(\cJ x)$ and expand it for large $\cJ$ at fixed $x$, getting $e^{\tint}(x)=e^{\tint}_{\treg}(x)+e^{\tint}_{\tsing}(x)$ where $e^{\tint}_{\tsing}$ is the part whose integral is divergent at $x=0$. 
     The regular part in one regime is in fact equal to the singular part in the other regime
     (as  in \adss  case this can be checked order by order in the large $\cJ$ expansion)
\beq
	e^{\tint}_{\tsing}(x)=e^{\tsum}_{\treg}(\cJ x), \ \ \ \ \ \ e^{\tsum}_{\tsing}(n)=e^{\tint}_{\treg}(n/\cJ),
\eeq
so that \rf{E1eHL} takes the form
\beqa
&&	\ \ \ \   E_1=E_1^{\rm analytic}+ E_1^{\rm non-analytic}  \ , \    \la{310} \\
&& 
	E_1^{\rm analytic}= \sum\limits_{n\in\mZ}^{} e^{\tsum}_{\treg}(n),\qquad  \qquad   E_1^{\rm non-analytic}\equiv 
		\delta E_1=\int\limits_{-\infty}^{\infty}\cJ dxe^{\tint}_{\treg}( x)\ .
\eeq
 $E_1^{\rm analytic}$  gives the ``{\it analytic}'' part of the 1-loop correction:
  its large $\cJ$ expansion contains only {\it even}
   powers of $\cJ$ which translate to {\it  integer} powers of the coupling $\lambda$ 
   if we rewrite the result in terms of the total 
   angular momentum $J=\rl\cJ$. 
   The   integral term, $\delta E_1$, gives the ``{\it non-analytic}'' contribution:
   it contains {\it odd } powers of $\cJ$ and thus leads to  {\it half-integer} powers of $\lambda$ when 
   expressed in terms of $J$. This non-analytic part was responsible in the $AdS_5\times S^5$ case for the famous ``3-loop discrepancy''.
    In the present  \adt  case we find  from \rf{E1eHL},\rf{eT4} 
\beq
\label{su2dET4}
	\delta E^{AdS_{3}}_1 = \frac{m^4}{2 \cJ^3}-\frac{7 m^6}{12 \cJ^5}+\frac{29 m^8}{48 \cJ^7}-\frac{97 m^{10}}{160
   \cJ^9}+\frac{2309 m^{12}}{3840 \cJ^{11}}+\dots\ .
\eeq
This expansion is reproduced by the following closed  expression
\beq
 \delta E^{AdS_{3}}_1=\frac{1}{\sqrt{\cJ^2+m^2}}\Big(m^2+\cJ^2 \log \frac{\cJ^2}{m^2+\cJ^2}\Big) \ , \la{314} 
\eeq
that we found  using the same  method  as used in  \cite{SchaferNameki:2006gk}, 
i.e.  by rewriting the sum as a contour integral in the $n$ plane. 
This method also shows that the analytic part is the same as it was for \adss case 
 because it is essentially determined by the $S^3$ frequencies \eq{su2S3omega} only.
  On the contrary, for the non-analytic part, the  contributions of {\it all}  frequencies are important.

For comparison, let us recall  the corresponding non-analytic  contribution in the  \adss case \ci{Beisert:2005cw,Minahan:2005qj}
\beqa
\nonumber
	\delta E^{AdS_{5}}_1
	&=&\frac{1}{\sqrt{\J^{2}+m^{2}}}\,\Big(
m^{2}+2\,\J^{2}\,\log\frac{\J^{2}}{\J^{2}+m^{2}}-\frac{\J^{2}-m^{2}}{2}\,\log\frac{\J^{2}-m^{2}}{\J^{2}+m^{2}}
\Big)\,\\ 
	\label{su2dEHL}
	&=&
	-\frac{m^6}{3 \cJ^5}+\frac{m^8}{3\cJ^7}-\frac{49 m^{10}}{120 \cJ^9}+\frac{2 m^{12}}{5
   \cJ^{11}}+\dots\ .
\eeqa

\subsection{Constraining the dressing phase in $SU(2)$ sector}

The above result for the non-analytic part of the one-loop energy is expected to originate from the dressing phase in the ABA equations.
The Bethe ansatz equations in the $su(2)$ sector, corresponding to 
strings with nontrivial motion only in the $S^3$ part of the background, are the same as in the \adss case:
\beq
\label{abasu2}
	\Big(\frac{x_{2,j}^{+}}{x_{2,j}^{-}}\Big)^{L} &=&
	\prod_{k\neq j}\frac{x_{2,j}^{+}-x_{2,k}^{-}}{x_{2,j}^{-}-x_{2,k}^{+}}
	\frac{1-\frac{1}{x_{2,j}^{+}x_{2,k}^{-}}}{1-\frac{1}{x_{2,j}^{-}x_{2,k}^{+}}}\,
	\sigma^{2}(x_{2,j}, x_{2,k})\ . 
\eeq
They are obtained from the full set of ABA equations in section 2  by considering states with only $x_2$ Bethe roots excited. 

From these equations we can compute the ABA prediction for the non-analytic part of the 1-loop string energy similarly to how this was 
 done in the \adss case. 
 Let us first assume that the dressing phase has the form \eq{BES} with summation starting from $r=2$ but keep coefficients $c_{r,s}^{(1)}$ unfixed. 
 Such structure of the phase would be in agreement with the proposal of \cite{Beisert:2005wv} 
 which is expected to apply to  a large class of integrable systems. Following the same method as used in 
 \cite{Hernandez:2006tk} we find that the Bethe ansatz then  predicts that the non-analytic part of the 1-loop energy should be 
\beq
\label{BAEa23}
	\delta E_1=\frac{m^6
   }{16 \cJ^5}c_{2,3}^{(1)}+\frac{m^8 }{64 \cJ^7}(-5 c_{2,3}^{(1)}-2 c_{2,5}^{(1)}+c_{3,4}^{(1)})	+\dots\ . 
\eeq
This prediction  is,  however,  in structural disagreement with the \adt  string 
result    in  \rf{su2dET4}  as the latter starts with a $\frac{1}{\cJ^3}$ term, i.e.
 one order earlier than the expansion \eq{su2dEHL} obtained for \adss  case.
 This difference may be attributed to the reduced amount of supersymmetry (and  thus 
 supersymmetry protection)    in the $AdS_3$ case  compared to $AdS_5$ one. 
 
 This forces us to modify the structure of \eq{BES}:
  we propose to include also the $c_{1,s}^{(1)}$ coefficients, i.e. to assume that   the  summation in  the phase   \eq{BES}   should 
  start  from $r=1$.
Then the Bethe ansatz prediction becomes
\beqa
\label{baesu2J3}
	\delta E_1&=&\frac{m^4 }{4 \cJ^3}c_{1,2}^{(1)}-\frac{m^6 }{16
   \cJ^5}(4 c_{1,2}^{(1)}+c_{1,4}^{(1)}-c_{2,3}^{(1)})\\ \nn
   & & \  +\ \frac{m^8 }{64
   \cJ^7}(15 c_{1,2}^{(1)}+5 c_{1,4}^{(1)}+2 c_{1,6}^{(1)}-5 c_{2,3}^{(1)}-2
   c_{2,5}^{(1)}+c_{3,4}^{(1)})+\dots \ ,  
\eeqa
which reduces to \rf{BAEa23}  in the  \adss   case   where one has $c_{1,s}^{(1)}=0$. 
Comparing to \rf{su2dET4}, we find that in the \adt case   
\beqa
\label{csu2f}
	c_{1,2}^{(1)}&=&2\ , \\ 
	c_{1,4}^{(1)}&=&\frac{4}{3}+c_{2,3}^{(1)}\ , \\ 
	\label{csu2l}
	c_{1, 6}^{(1)} &=& \frac{1}{2} (2 c_{2, 5}^{(1)} - c_{3, 4}^{(1)} + 2) \ . 
\eeqa
Our proposed coefficients $c_{r,s}^{(1)}$   in \rf{new} are  consistent with these relations.

In the above    derivation 
we assumed that the 1-loop correction to $h(\lambda)$ is zero, i.e.  
\beq
	h(\lambda)=\frac{\sql}{4\pi}+ \c +\O\Big(\frac{1}{\rl}\Big)  \  , \ \ \ \ \ \ \ \ \ \   \c_{_{\rm AdS_3 \times S^3 \times T^4}}=0 \ . \la{322}
\eeq
 Having $\c$ nonzero would produce (after replacing $\sql$  by $4 \pi h$ in the classical energy  \rf{33}) 
   an extra $\frac{1}{\cJ}$ term  in   $\delta E^{AdS_3}_1$  which is absent in \rf{su2dET4}. 

Let us note that the {\it analytic} part of the energy, which,  as we discussed in Section 3,  is the same in \adt and \adss  cases, is correctly
 reproduced from the ABA \eq{abasu2}, since it is only sensitive to the AFS part of the phase.

The dressing phase  should   be universal,  i.e.  the same    phase should   be possible  to extract also 
from the study of other classical  solutions.  Indeed, as we shall  find in Appendix A, 
the same relations \rf{csu2f}--\rf{csu2l}   follow also from the   expression for the non-analytic part of  the 1-loop 
energy of the   large spin $(S,J)$ folded string solution  in the $SL(2)$ sector. This is not totally surprising as the two solutions  are 
related by an analytic continuation \cite{Frolov:2006qe}; nevertheless, 
this is a nontrivial check, since on the Bethe ansatz side the two calculations are quite different. 
Moreover, as we shall   explain in detail  in sections 4-6, the same  1-loop   phase  can be found    for a generic 
semiclassical solution using the algebraic curve method  used in \adss  case in \ci{Gromov:2007cd}.

\renewcommand{\theequation}{4.\arabic{equation}}
\setcounter{equation}{0}
\section{Semiclassical dressing  phase from the algebraic curve approach}
\label{semicl}


Considering the  strong-coupling (string semiclassical) expansion in the Bethe equations, 
the leading term   is   given by the  integral   equations parametrized by an algebraic curve which 
represents a   generic  finite gap string solution. Starting with an algebraic curve description of such  string solution one may 
compute the 1-loop  correction   by summing up the corresponding fluctuation frequences 
and then extract the dressing phase contribution. 
This powerful approach has been developed in 
\cite{Gromov:2007cd,Gromov:2007aq} for the $AdS_{5}\times S^{5}$ case (see also \cite{SchaferNameki:2010jy} where the algebraic curve method is reviewed). Here  we  will use the same method   in the 
$AdS_{3}\times S^{3}\times T^{4}$  case.

\subsection{Scaling limit of the Bethe equations and finite-gap equations}

Let us introduce the function\footnote{We use $\wh\alpha(x)$   notation 
 instead of the    standard  $\alpha(x)$ to avoid confusion with the background parameter $\alpha$  used  in other sections.}
\be
\wh\alpha(x) = \frac{4\,\pi}{\sqrt\lambda}\,\frac{x^{2}}{x^{2}-1}\ ,
\ee
and define the discrete resolvents
\be
\label{eq:GH-defs}
G_{a}(x) = \sum_{k=1}^{K_{a}} \frac{\wh \alpha(x_{a,k})}{x-x_{a,k}},\qquad
H_{a}(x) = \sum_{k=1}^{K_{a}} \frac{\wh \alpha(x)}{x-x_{a,k}},\qquad
\overline G(x) = G(1/x), \quad \overline H(x) = H(1/x).
\ee
Expand the quantum Bethe equations equations at large $h\sim \sql $
we find for   first three   equations in  \rf{mi}--\rf{mii} 
\ba
&& x\in \mathcal C_{1}, \qquad 2\,\pi\,n_{1} = -G_{2}-\overline H_{\overline 2}-\frac{G_{\overline 2}(0)+G'_{\overline 2}(0)\,x}{x^{2}-1}, \no \\
&& x\in\mathcal C_{2}, \qquad 2\,\pi\,n_{2} +\frac{4\,\pi\,\J\,x}{x^{2}-1} = 2\,H_{2}-H_{1}-H_{3}+
2\,\overline H_{\overline 2}-\overline H_{\overline 1}-\overline H_{\overline 3}+2\,
\frac{G_{\overline 2}(0)-G_{2}(0)}{x^{2}-1}, \nonumber \\
&& x\in \mathcal C_{3}, \qquad 2\,\pi\,n_{3} = -G_{2}-\overline H_{\overline 2}-\frac{G_{\overline 2}(0)+G'_{\overline 2}(0)\,x}{x^{2}-1}.\la{43}
\ea
These match the finite-gap equations in Eqs.(7.39)-(7.41) of \cite{Babichenko:2009dk} upon use of the identity
\be
G(x) = H(x)-\frac{G(0)+G'(0)\,x}{x^{2}-1}.
\ee

\subsection{Quasi-momenta and algebraic curve}

The finite-gap equations can be written as (the equation with $n_{\ell}$ is evaluated at $x\in \mathcal C_{\ell}$)
\ba
&& 2\,\pi\,n_{1} = -H_{2}-\overline H_{\overline 2}+\frac{G_{2}(0)+x\,G_{2}'(0)}{x^{2}-1}
-\frac{G_{\overline 2}(0)+x\,G_{\overline 2}'(0)}{x^{2}-1}, \\
&& 2\,\pi\,n_{2}+\frac{4\,\pi\,\J\,x}{x^{2}-1} = 2\,H_{2}-H_{1}-H_{3}+2\,\overline H_{\overline 2}-
\overline H_{\overline 1}-\overline H_{\overline 3}+2\,\frac{G_{\overline 2}(0)-G_{2}(0)}{x^{2}-1}, \\
&& 2\,\pi\,n_{3} = -H_{2}-\overline H_{\overline 2}+\frac{G_{2}(0)+x\,G_{2}'(0)}{x^{2}-1}
-\frac{G_{\overline 2}(0)+x\,G_{\overline 2}'(0)}{x^{2}-1}, \\
&& 2\,\pi\,n_{\overline 1} = -H_{\overline 2}-\overline H_{2}-\frac{G_{2}(0)+x\,G_{2}'(0)}{x^{2}-1}
+\frac{G_{\overline 2}(0)+x\,G_{\overline 2}'(0)}{x^{2}-1}, \\
&& 2\,\pi\,n_{\overline 2}-\frac{4\,\pi\,\J\,x}{x^{2}-1} = 2\,H_{\overline 2}-H_{\overline 1}-H_{\overline 3}+2\,\overline H_{ 2}-
\overline H_{ 1}-\overline H_{ 3}-2\,\frac{G_{\overline 2}(0)-G_{2}(0)}{x^{2}-1}, \\
&& 2\,\pi\,n_{\overline 3} = -H_{\overline 2}-\overline H_{ 2}-\frac{G_{2}(0)+x\,G_{2}'(0)}{x^{2}-1}
+\frac{G_{\overline 2}(0)+x\,G_{\overline 2}'(0)}{x^{2}-1}.
\ea
Building on the work of \cite{Beisert:2005bm}, we set~\footnote{For bosonic classical solutions in 
$PSU(1,1|2)$ we can set $p_{1}+p_{4}=p_{2}+p_{3}=0$ and similarly for the barred quasi-momenta.}
\ba
p_{1}  =  -p_{4} &=& -\frac{1}{2}\,H_{1}-\frac{1}{2}\,\overline H_{\overline 1}-\frac{1}{2}\,H_{3}-
\frac{1}{2}\,\overline H_{\overline 3}-\frac{2\,\pi\,\J\,x}{x^{2}-1}+\frac{x}{x^{2}-1}\big[
G_{2}'(0)-G_{\overline 2}'(0)
\big], \\
p_{2}  =  -p_{3} &=& H_{2}+\overline H_{\overline 2}-\frac{1}{2}\,H_{1}-\frac{1}{2}\,\overline H_{\overline 1}-\frac{1}{2}\,H_{3}-
\frac{1}{2}\,\overline H_{\overline 3}-\frac{2\,\pi\,\J\,x}{x^{2}-1}, \\
p_{\overline 1}  =  -p_{\overline 4} &=& -\frac{1}{2}\,H_{\overline 1}-\frac{1}{2}\,\overline H_{ 1}-\frac{1}{2}\,H_{\overline 3}-
\frac{1}{2}\,\overline H_{ 3}+\frac{2\,\pi\,\J\,x}{x^{2}-1}+\frac{x}{x^{2}-1}\big[
G_{\overline 2}'(0)-G_{2}'(0)
\big], \\
p_{\overline 2}  =  -p_{\overline 3} &=& H_{\overline 2}+\overline H_{ 2}-\frac{1}{2}\,H_{\overline 1}-\frac{1}{2}\,\overline H_{ 1}-\frac{1}{2}\,H_{\overline 3}-
\frac{1}{2}\,\overline H_{ 3}+\frac{2\,\pi\,\J\,x}{x^{2}-1}.
\ea
Up to winding contributions, we have 
\be
p_{1,2,3,4}(x) = p_{\overline 1, \overline 2, \overline 3, \overline 4}(1/x)\ . 
\ee
The above finite-gap equations are obtained with $p_{i}-p_{j} = 2\,\pi\,n_{ij}$ where 
\be
(i,j) = (1,2), (2,3), (3,4), \qquad (\overline 1, \overline 2), (\overline 2, \overline 3), (\overline 3, \overline 4).
\ee 
Note that here the algebraic curve is a connected sum of two pieces interchanged by the $x\to 1/x$ transformation, while 
in the  $AdS_{5}\times S^{5}$  case the curve is a single connected invariant piece.

\subsection{Semiclassical one-loop dressing factor}

The semiclassical one-loop
dressing factor is built according to the prescription in \cite{Gromov:2007cd}.  Fig.1  
gives  the picture of the Dynkin nodes, algebraic curve sheets and physical fluctuations for the unbarred $PSU(1,1|2)$ factor.

\begin{figure}[H]
\begin{center}
\includegraphics[scale=1]{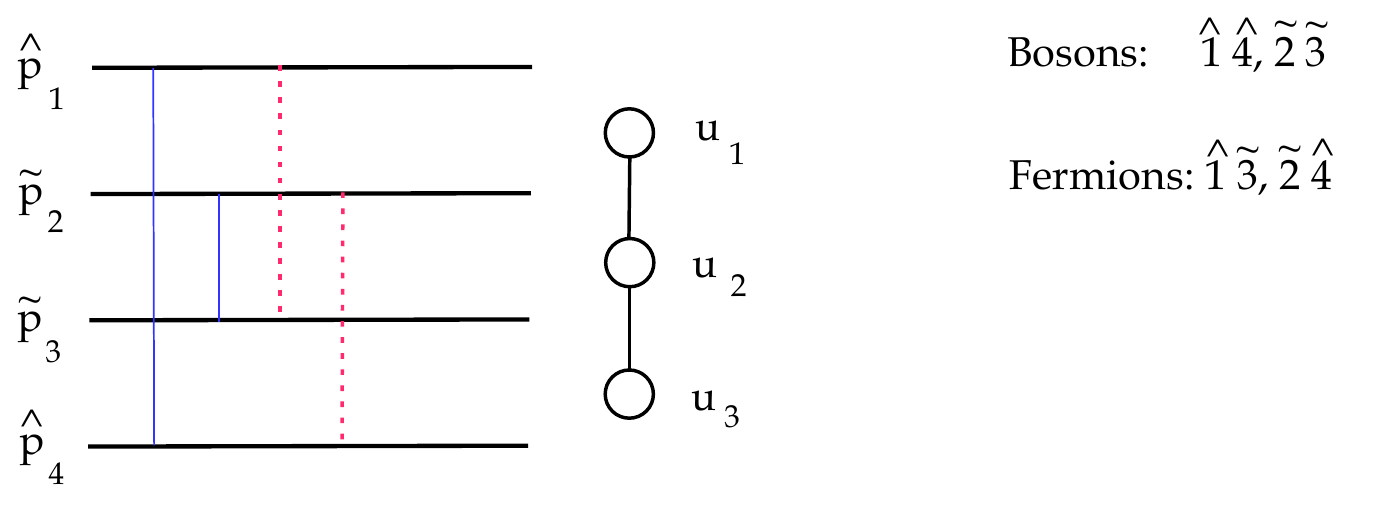}
\caption{Setup for computation of $\mathcal V$.} 
\label{fig:dressing}
\end{center}
\end{figure}

 For each quasi-momentum $p_I$ (with $I\in\{1,2,3,4,\bar1,\bar2,\bar3,\bar4\}$),  we are to  compute 
the corresponding correction $V_{I}$ which is the following sum over all polarizations (i.e.  from both $PSU(1,1|2)$ symmetry factors)
\be
V_{I}(x) = \frac{1}{2}\sum_{ij}(-1)^{F_{ij}}\int \frac{dy}{2\,\pi}\,(p_{i}'-p_{j}')\,\Big[
\delta H_{I}^{ij}\,\frac{\wh \alpha(x)}{x-y}+\delta \overline H_{I}^{ij}\,\frac{\wh \alpha(1/x)}{1/x-y}
\Big].
\ee
Here, for each polarization $(ij)$, we determine the Bethe roots $\mathcal B_{ij}\subset \{u_{1, 2,3}, u_{\overline 1, \overline 2, \overline 3}\}$ that 
 lie between the sheets $i$ and $j$ and 
evaluate the total variation of the functions $H$ and $\overline H$ appearing in $p_{I}$ due to the addition of one root
for each element of $\mathcal B_{ij}$. The total phase corrections to the Bethe equations are obtained by evaluating $V_{I}-V_{J}$. The result for the middle node 2 equation is given by $\mathcal V(x)\equiv V_2(x)-V_3(x)$ and reads (we follow the notation of \cite{Gromov:2007cd} for the integral)
\be
\label{eq:complete-V2}
\mathcal V(x) = \int_{-1}^{1}\frac{dy}{2\,\pi}\,\Big[\Big(G_{2}(y)+\overline G_{\overline 2}(y)\Big)'\,\frac{\wh \alpha(x)}{x-y}
+\Big(\overline G_{2}(y)+ G_{\overline 2}(y)\Big)'\,\frac{\wh \alpha(1/x)}{1/x-y}\Big].
\ee
The equations for the nodes 1, 3 are not corrected. 
The potential $\mathcal V$ contains the correction to the two dressing contributions. If we excite only the node 2, it reduces to 
\be
\label{eq:V22}
\mathcal V_{2}(x) =\int_{-1}^{1}\frac{dy}{2\,\pi}\,\Big[G_{2}'(y)\,\frac{\wh \alpha(x)}{x-y}
+\overline G_{2}'(y)\,\frac{\wh \alpha(1/x)}{1/x-y}\Big],
\ee
where the notation is $\int_{-1}^{1} = \frac{1}{2}\int_{C^{+}}+\frac{1}{2}\int_{C^{-}}$ and the half circumferences
$C^{\pm}$ (and their orientation) are defined in the caption of figure 4 of \cite{Gromov:2007cd}.

\def \wc {{\wh c}}

The next step is to evaluate $\mathcal V_{2}(x)$ for large $x$. This is done by factoring  out $\wh\alpha(x)$ in the 
integrand of (\ref{eq:V22}) and expanding at large $x$. Using the relation between $G_{2}$ and the charges $Q_{n}$,
\be
G_{2}(y) = -\sum_{n=0}^{\infty} Q_{n+1}\,y^{n},
\ee
the resulting function of $y$ is not singular anywhere on the 
circle $|y|=1$ and  the integration is trivial. The result is
\ba
\label{eq:V2}
&&\mathcal V_{2}(x) = \frac{\wh \alpha(x)}{2\,\pi}\,\sum_{r=2}^{\infty}\sum_{s=1}^{\infty} \wc_{r,s}\,\frac{Q_{r}}{x^{s}}\ , \\
\label{wcvals}
&& \wc_{r,s} = -4\,\Big(1-\frac{1}{2}\delta_{s,1}\Big)\,\frac{1-(-1)^{r+s}}{2}\,\frac{r-1}{r+s-2}\,\ .   
 \ee
Here  $\wc_{r,s}$ are the (naive) 
prediction of the algebraic curve method for the values of the  $c_{r,s}^{(1)}$ coefficients which parametrize the phase according to \eq{cexp}. We  notice that the coefficients $\wc_{r,s}$  are not antisymmetric. 
This is  
 a serious   problem   since  the   
 antisymmetry of the coefficients $c_{r,s}$ in the phase 
  is an important  consistency requirement 
  (see section 4.4 below).

 Indeed, 
in  section 5 we  will 
 show on the example of the $SU(2)$ sector circular string 
 that the part that breaks the antisymmetry induces a mismatch   with the string  theory result \rf{314} for 
  the non-analytic  term   in the 
 one-loop energy.   This  disagreement turns out to be   due to a regularization ambiguity 
   in the sum over fluctuation frequencies.   
 Once this regularization problem is fixed, the  algebraic curve approach result   agrees   with the string  theory result  and the 
  the antisymmetry of $c_{r,s}$   is recovered. 

\subsection{On  consistency condition on the phase}

Let us make a comment  concerning the orgin of  antisymmetry of the phase coefficients
$c_{r,s}$    and why this antisymmetry is  {\it not}  obvious  
in the   algebraic curve approach  in the present $AdS_3$ case.

 If we consider the $\mathfrak{sl}(2)$ Bethe equations written in the form 
\be
\Big(\frac{x^{+}_{i}}{x^{-}_{i}}\Big)^{J} = \prod_{j\neq i}
\frac{x_{i}^{-}-x_{j}^{+}}{x_{i}^{+}-x_{j}^{-}}\,
\frac{1-\frac{1}{x_{i}^{+}\,x_{j}^{-}}}{1-\frac{1}{x_{i}^{-}\,x_{j}^{+}}}\,e^{i\,\vartheta_{ij}} 
\ee
and take the product over $i$, we find 
\be
\label{sumij}
\sum_{i,j}\vartheta_{ij} = 0.
\ee
This is automatic if the coefficients $c_{r,s}$ which define the phase $\vartheta_{ij}$ are antisymmetric. 

While  this constraint  is  thus a direct  consequence of the discrete form of ABA  equations, it is not automatic in the 
thermodynamic  (``semiclassical'') limit due to  infinite  summation and thus regularization issues involved. 
The relation \eq{sumij} implies  the following condition for the potential 
\rf{eq:complete-V2}  that determines the dressing phase:
\be
\sum_{k} \mathcal V(x_{k}) = 0 \ .
\ee
In  the  $AdS_{5}\times S^5$  case  we have 
\be
\mathcal V^{AdS_{5}}(x) = \int_{-1}^{1}\frac{dy}{2\pi}\Big[G(y)-G(1/y)\Big]'\,\Big[\frac{\wh\alpha(x)}{x-y}-\frac{\wh\alpha(1/x)}{1/x-y}\Big]\,.
\ee
Using the relations 
\be
\sum_{k} \frac{\wh\alpha(x_{k})}{x_{k}-y} = -G(y), \qquad \qquad 
\sum_{k} \frac{\wh\alpha(1/x_{k})}{1/x_{k}-y} = G(0)-G(1/y), 
\ee
we find indeed that 
\be
\sum_{k} \mathcal V^{AdS_{5}}(x_{k}) = \int_{-1}^{1}\frac{dy}{2\pi}\Big[
-\frac{1}{2}\Big(G(y)-G(1/y)\Big)^{2}-G(0)\Big(G(y)-G(1/y)\Big)
\Big]' = 0\ .
\ee
Instead,  in  the $AdS_{3}\times S^3 \times T^4$  case  we have 
\be
\mathcal V^{AdS_{3}}(x) = \int_{-1}^{1}\frac{dy}{2\pi}\Big[
G'(y)\,\frac{\wh\alpha(x)}{x-y}+G'(1/y)\,\frac{\wh\alpha(1/x)}{1/x-y}
\Big]\,,
\ee
and thus 
\be
\sum_{k} \mathcal V^{AdS_{3}}(x_{k}) = \int_{-1}^{1}\frac{dy}{2\pi}\Big\{-\frac{1}{2}\Big[G^{2}(y)+G^{2}(1/y)\Big]
+G(0)G(1/y)\Big\}'
\ee
does  not vanish automatically. 
Even  assuming  that  
$G(0)=0$  we get  an a priori  non-vanishing term 
\be
\sum_{k} \mathcal V^{AdS_{3}}(x_{k}) = -\frac{1}{2\pi}\,\Big[ G^2(1)-G^2(-1)\Big]\ ,\la{vet}
\ee
implying  that the dressing phase coefficients $c_{r,s}$ coming from this potential will not be automatically antisymmetric.
This is, indeed,  what we have  found above in \rf{wcvals}.

However,  as we shall   explain  below, it is possible  to adjust the regularization involved in the definition of the ``semiclassical'' limit 
 (subtracting from the potential a regularization-related part)
  so that   to ensure  the vanishing  of \rf{vet}  and thus  the   antisymmetry of the $c_{r,s}$ coefficients in  the \adt theory.


\renewcommand{\theequation}{5.\arabic{equation}}
\setcounter{equation}{0}

\section{Algebraic curve  approach  applied to $SU(2)$ circular string case:  regularization ambiguity}

Let   us now apply the discussion of the previous section  to the circular string example 
discussed   from the string theory  perspective in section 3.

\def \DE  {{\Delta E_1}}

\subsection{Non-analytic  part of  one-loop energy}

The expansion of $\mathcal V_{2}(x)$  in \rf{eq:V22},\rf{eq:V2}  
can   be analysed   using  the strategy developed in \cite{Hernandez:2006tk}. One
considers a dressing contribution which is the usual combination \rf{BES} 
of charges  with some unknown coefficients $c_{r,s}$. Then
one perturbs the quadratic equation for the 
resolvent associated to a given  solution of the finite-gap integral equation. The result is a compact expression for the 
dressing correction to the string energy. 
Applied to the $SU(2)$ circular  string case 
the ingredients in this expression 
 are the classical charges $Q_{n}(m, \cJ) $ defined by 
\be
-\sum_{n=0}^{\infty} Q_{n+1}\,x^{n} = 2\,\pi\,m+\frac{\sqrt{1+(m/\J )^{2}}-\sqrt{1+(4\pi m x)^{2}}}
{2\,\big(x-\frac{1}{(4\,\pi\,\J)^{2}}\,\frac{1}{x})}.
\ee
The energy correction for the perturbation associated with  $\mathcal V_{2}$ 
computed in the algebraic curve approach is then 
\be
\delta E^{AC}_1= -\frac{1}{\pi\,\big[1+\frac{2\,Q_{2}}{(4\,\pi\,\J)^{2}}\big]}\mathop{\sum_{r\ge 2}}_{s\ge 1} 
\Big(\frac{1}{4\,\pi\,\J}\Big)^{r+s}\,\wc_{r,s}\,Q_{s+1}\,Q_{r}. \la{acc}
\ee
where  the coefficients $\wc_{r,s}$ were defined in \eq{wcvals}.
Expanding this correction at large $\J$ we get
\be
\delta E^{AC}_1 = \frac{m^4}{2 \mathcal{J}^3}-\frac{17 m^6}{24 \mathcal{J}^5}+\frac{41 m^8}{48
   \mathcal{J}^7}-\frac{623 m^{10}}{640 \mathcal{J}^9}+\frac{4139 m^{12}}{3840
   \mathcal{J}^{11}}-\frac{126079 m^{14}}{107520 \mathcal{J}^{13}}+\frac{18069 m^{16}}{14336
   \mathcal{J}^{15}}+\dots.
  \ee
  This is nothing but the expansion of 
\be
\label{mis}
\delta E^{AC}_1 = \underbrace{
\frac{1}{\sqrt{m^{2}+\J^{2}}}\Big(
m^{2}+\J^{2}\,\log\frac{\J^{2}}{m^{2}+\J^{2}}\Big)}_{\delta E_1^{AdS_{3}}}-\underbrace{
\frac{m^2 \left(2 \mathcal{J} \left(\mathcal{J}-\sqrt{\mathcal{J}^2+m^2}\right)+m^2\right)}{2
   \left(\mathcal{J}^2+m^2\right)^{3/2}}}_{\DE }\  , 
\ee
where $\delta E^{AdS_{3}}_1$ is the string-theory  result  in  (\ref{314}), while $\DE$
is a discrepancy.  As we shall explain below, the latter   is  related to  an implicit choice of regularization 
in the  
 algebraic curve approach.\footnote{For the $AdS_4\times\mathbb{CP}^3$ background, related regularization issues were discussed  in \cite{Gromov:2008fy} for the folded string and in \cite{Abbott:2010yb} for giant magnon solutions.
}

\subsection{Regularization origin of the mismatch}

Let us  rederive the result \rf{mis}  directly.
First, let us relabel the quasi momenta for the two  $PSU(1,1|2)$ factors as follows
\be
\underbrace{\wh p_{1}, \wt p_{2}, \wt p_{3}, \wh p_{4}}_{PSU(1,1|2)_{+}}, \qquad
\underbrace{\wh p_{2}, \wt p_{1}, \wt p_{4}, \wh p_{3}}_{PSU(1,1|2)_{-}}, \qquad
\ee
The physical $4_{B}+4_{F}$ (massive) polarizations are organized as
\ba
B&:&\ \  (\wt 2\ \wt 3 ), \ (\wh 1\ \wh 4), \ (\wt 1\ \wt 4), \ (\wh 2\ \wh 3)\\
F&:&\ \  (\wh 1\ \wt 3 ), \ (\wt 2\ \wh 4), \ (\wh 2\ \wt 4), \ (\wt 1\ \wh3).
\ea
The explicit quasi momenta are ($\kappa=\sqrt{\J^{2}+m^{2}}$)
\ba
\wh p_{1} &=& \wh p_{2} = -\wh p_{3} = -\wh p_{4} = \frac{2\,\pi\,\kappa\,x}{x^{2}-1}, \\
\wt p_{2} &=& -\wt p_{3} = 2\,\pi\,\frac{x}{x^{2}-1}\,\sqrt{m^{2}\,x^{2}+\J^{2}}-2\,\pi\,m, \\
\wt p_{1} &=& -\wt p_{4} = 2\,\pi\,\frac{x}{x^{2}-1}\,\sqrt{m^{2}/x^{2}+\J^{2}}.
\ea
Computing the off-shell frequencies $\Omega^{ij}$ for each polarization and defining
\be
\Omega^{ij}_{n} = \Omega^{ij}(x^{ij}_{n}), 
\ee
where $x^{ij}_{n}$ is the solution of $p_{i}-p_{j} = 2\,\pi\,n$, we find 
\ba
\Omega^{\wt 2\ \wt 3}_{n} &=& \sqrt{\frac{2\J^{2}+M^{2}-2\sqrt{\J^{4}+m^{2}M^{2}+\J^{2}M^{2}}}{\J^{2}+m^{2}}}, \qquad M = n+2m, \\
\Omega^{\wh 1\ \wh 4}_{n} &=& \sqrt\frac{\J^{2}+m^{2}+n^{2}}{\J^{2}+m^{2}}-1, \\
\Omega^{\wh 1\ \wt 3}_{n} &=& \Omega^{\wt 2\ \wh 4}_{n} = \sqrt\frac{\J^{2}+M^{2}}{\J^{2}+m^{2}}-1,
\qquad M = n+m, \\
\Omega^{\wt 1\ \wt 4}_{n} &=& -\frac{2\J}{\kappa}+\sqrt{\frac{2\J^{2}+M^{2}+2\sqrt{\J^{4}+m^{2}M^{2}+\J^{2}M^{2}}}{\J^{2}+m^{2}}}, \qquad M = n, \\
\Omega^{\wh 2\ \wh 3}_{n} &=& \sqrt\frac{\J^{2}+m^{2}+n^{2}}{\J^{2}+m^{2}}-1, \\
\Omega^{\wh 2\ \wt 4}_{n} &=& \Omega^{\wt 1\ \wh 3}_{n} = -\frac{\J}{\kappa}+\sqrt\frac{\J^{2}+M^{2}}{\J^{2}+m^{2}}-1,\qquad M = n.
\ea
Notice that for the evaluation of the non-analytic part of the   1-loop   correction to  the energy, 
any finite set of modes is irrelevant since    each mode  contribution is separately analytic in the large $\J$ expansion.
 The non-analytic contribution arises from the infinite summation. 
 Note also that, with our choice of quasimomenta, the first set of four frequencies have
  a shift $M\neq n$. In terms of $M$, these frequencies are the same as in the  string world-sheet calculation in section 3.

The one-loop energy can then  be computed as usual  as the  sum over polarizations  or in terms of the   integral 
representation  (see, e.g.,  \ci{Gromov:2007cd,Gromov:2007aq,Gromov:2008ec}) 
\be
E_1 = \sum_{ij}(-1)^{F_{ij}} \oint \frac{dx}{2\,\pi\,i}\, \Omega^{ij}(x) \, \partial_{x} \log\sin\frac{p_{i}-p_{j}}{2\,\pi}, \la{gr} 
\ee
where the integration encircles the points  $x_{n}^{ij}$. This contour can be transformed  in the unit circumference
as usual up to cut terms that do not contribute to the non-analytic part.  Neglecting exponentially suppressed contributions 
in the large $\J$ limit, we checked numerically that from \rf{gr} 
we obtain precisely the result (\ref{mis}), i.e. the sum of  
$\delta E^{AdS_{3}}_1$ plus an extra term $\DE$.

\begin{figure}[H]
\begin{center}
\includegraphics[scale=0.9]{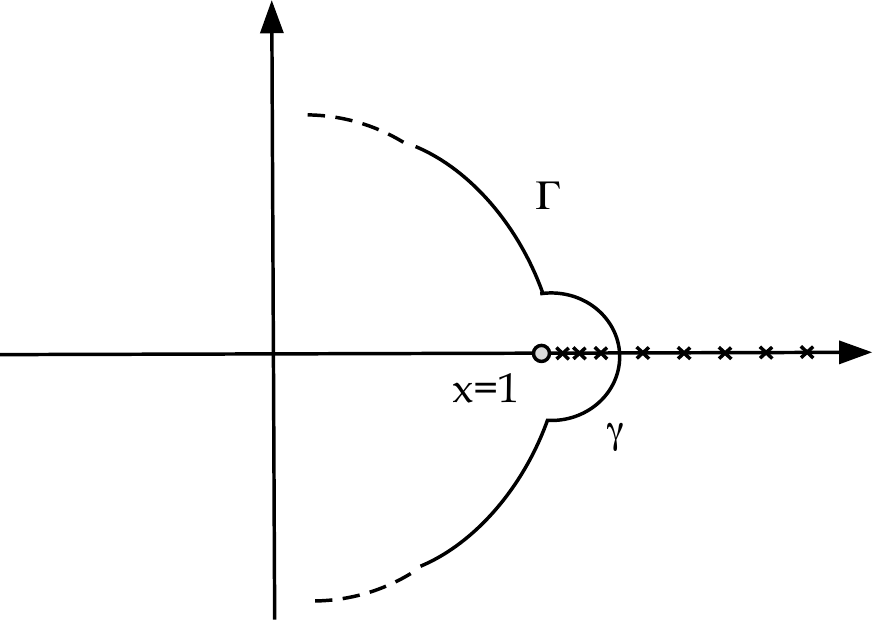}
\caption{Contour defining the AC regularization} 
\label{fig:contour}
\end{center}
\end{figure}

Let  now show that the origin of the extra term $\DE$   is due to a particular choice of 
 regularization used in the algebraic curve approach.
When we evaluate the unit circumference contribution, we consider a contour like that shown in Fig.~(\ref{fig:contour})
and it is the same for all polarizations.

The crosses are the poles $x_{n}^{ij}$ and the small part of circle $\gamma$ around $x=1$ determines a cut-off 
on $n$ that depends on the polarization $(ij)$. To see this, one can expand at small $\varepsilon$ the differences
\be M_{ij} = { 1 \ov 2 \pi} \big[ p_{i}(x)-p_{j}(x)\big]\ee
 after setting~\footnote{Here  the expression for  $x$ in terms of $\epsilon$ is {\it not} an expansion, but just a convenient quadratic parametrization $x(\varepsilon)$ with the property that $x(\varepsilon)\to 1$ when $\varepsilon\to 0$.}
\be
x = 1+\kappa\,\varepsilon+\frac{\kappa^{2}}{2}\,\varepsilon^{2}.
\ee
The explicit results are
\be
M_{\wt 2\ \wt 3} &=& \frac{1}{\epsilon }+\Big(\frac{m^2}{\sqrt{m^2+\mathcal{J}^2}}-2 m\Big)+\Big(\frac{m^4}{2 m^2+2 \mathcal{J}^2}+\frac{2 \mathcal{J}^2 m^2}{2 m^2+2 \mathcal{J}^2}\Big) \epsilon +O\left(\epsilon ^2\right), \\
M_{\wh 1\ \wh 4} &=& \frac{1}{\epsilon }+O\left(\epsilon ^2\right), \\
M_{\wh 1\ \wt 3} &=& \frac{1}{\epsilon }+\Big(\frac{m^2}{2 \sqrt{m^2+\mathcal{J}^2}}-m\Big)+\Big(\frac{m^4}{4 m^2+4 \mathcal{J}^2}+\frac{2 \mathcal{J}^2 m^2}{4 m^2+4 \mathcal{J}^2}\Big) \epsilon +O\left(\epsilon ^2\right),\\
M_{\wt 2\ \wh 4} &=& \frac{1}{\epsilon }+\Big(\frac{m^2}{2 \sqrt{m^2+\mathcal{J}^2}}-m\Big)+\Big(\frac{m^4}{4 m^2+4 \mathcal{J}^2}+\frac{2 \mathcal{J}^2 m^2}{4 m^2+4 \mathcal{J}^2}\Big) \epsilon +O\left(\epsilon ^2\right),\\
M_{\wt 1\ \wt 4} &=& \frac{1}{\epsilon }-\frac{m^2}{\sqrt{m^2+\mathcal{J}^2}}+\Big(\frac{m^4}{2 m^2+2 \mathcal{J}^2}+\frac{2 \mathcal{J}^2 m^2}{2 m^2+2 \mathcal{J}^2}\Big) \epsilon +O\left(\epsilon ^2\right),\\
M_{\wh 2\ \wh 3} &=& \frac{1}{\epsilon }+O\left(\epsilon ^2\right),\\
M_{\wh 2\ \wt 4} &=& \frac{1}{\epsilon }-\frac{m^2}{2 \sqrt{m^2+\mathcal{J}^2}}+\Big(\frac{m^4}{4 m^2+4 \mathcal{J}^2}+\frac{2 \mathcal{J}^2 m^2}{4 m^2+4 \mathcal{J}^2}\Big) \epsilon +O\left(\epsilon ^2\right)\\
M_{\wt 1\ \wh 3} &=& \frac{1}{\epsilon }-\frac{m^2}{2 \sqrt{m^2+\mathcal{J}^2}}+\Big(\frac{m^4}{4 m^2+4 \mathcal{J}^2}+\frac{2 \mathcal{J}^2 m^2}{4 m^2+4 \mathcal{J}^2}\Big) \epsilon +O\left(\epsilon ^2\right).
\ea
At this point, it is convenient to introduce the ``offsets''
\be
\Delta_{ij} = (-2m, 0, -m, -m, 0,0,0,0). 
\ee
These are such that the sum over the frequencies with a common (i.e. {\it universal}) 
 cutoff $M \le \frac{1}{\varepsilon}+\Delta_{ij}$
reproduces by construction  the string-theory   result $\delta E^{AdS_{3}}_1$  in the $\varepsilon\to 0$ limit. 
This  may be called  the {\em standard regularization}  for the non-analytic (dressing)  contribution, 
as compared to the  {\em AC regularization}  which is 
 $M\le M_{ij}$ with $\varepsilon\to 0$. 

To evaluate the  
difference between the results for  the  two regularization
prescriptions 
 we have to evaluate the additional terms in the sum from $M=\frac{1}{\varepsilon}+\Delta_{ij}$ to $M=M_{ij}$.
 Using the  Maclaurin summation   formula we get 
\be
\DE = \lim_{\varepsilon\to 0}\sum_{ij}(-1)^{F_{ij}}\Big[
\frac{1}{2}\Omega^{ij}\Big(\frac{1}{\varepsilon}+\Delta_{ij}\Big)+
\int_{\frac{1}{\varepsilon}+\Delta_{ij}}^{M_{ij}}dM\,\Omega^{ij}(M)\Big].
\ee
The integral  here can be done by replacing the integrand by its $\mathcal O(x)$ and $\mathcal O(x^{0})$ terms in the 
large $x$ expansion up to terms that vanish as $\varepsilon\to 0$. The computation gives exactly the expression in \rf{mis}
\be
\DE = -\frac{m^2 \Big[2 \mathcal{J} \left( \mathcal{J}-\sqrt{\mathcal{J}^2+m^2}\right)+m^2\Big]}{2
   \left(\mathcal{J}^2+m^2\right)^{3/2}} \ ,\la{de}
\ee
explaining the regularization origin of the discrepancy. 

\renewcommand{\theequation}{6.\arabic{equation}}
\setcounter{equation}{0}
\section{Proposal   for the  1-loop  dressing phase  coefficients from  the \\
algebraic curve approach}

As was   noted in section 4,  
 one should expect to find a  set of antisymmetric coefficients $c_{r,s}$  in $\vartheta$ in \rf{BES} 
 as this is  a consequence of the 
antisymmetry of the elementary magnon scattering phases.
The regularization  ambiguity that 
we have discussed  in the previous section  should be fixed  to ensure this antisymmetry. 
 Here   we  will  show that enforcing the  antisymmetry of  $c_{r,s}$    selects, for the $SU(2)$ circular string case discussed above, 
    the {\it standard} 
 regularization, removing the mismatch with string-theory result. 

Motivated by a  discussion in   \cite{David:2010yg}  let us  try to enforce the antisymmetry by integrating by parts.
We start with the 
 potential correcting the 
 Bethe equation  for the left moving sector middle node  (\ref{eq:complete-V2}), i.e.  (here $f' = \frac{\partial}{\partial y}f$)
\be
\mathcal V(x) = \int_{-1}^{1}\frac{dy}{2\,\pi}\,\Big[\Big(G_{2}(y)+\overline G_{\overline 2}(y)\Big)'\,\frac{\wh \alpha(x)}{x-y}
+\Big(\overline G_{2}(y)+ G_{\overline 2}(y)\Big)'\,\frac{\wh \alpha(1/x)}{1/x-y}\Big].
\ee
Let us integrate by parts and define 
\be
\widehat{\mathcal{V}}(x) =  \int_{-1}^{1}\frac{dy}{2\,\pi}\Big[\Big(G_{2}(y)+\overline G_{\overline 2}(y)\Big)'\,\frac{\wh \alpha(x)}{x-y}
-\Big(\overline G_{2}(y)+ G_{\overline 2}(y)\Big)\,\Big(\frac{\wh \alpha(1/x)}{1/x-y}\Big)'\Big].  \la{core} 
\ee
The difference is 
\be
\label{eq:diff}
\mathcal V(x)-\widehat{\mathcal V}(x) = \frac{\wh \alpha(1/x)}{2\,\pi}\,\Big[
\frac{G_{2}(1)+G_{\overline 2}(1)}{1/x-1}-\frac{G_{2}(-1)+G_{\overline 2}(-1)}{1/x+1}\Big].
\ee
The large $x$ expansion of $\widehat{\mathcal V}(x)$ is 
\be
\widehat{\mathcal V}(x) = \frac{\wh \alpha(x)}{2\,\pi}\,\sum_{r=1}^{\infty}\sum_{s=1}^{\infty}
\frac{c_{r,s}^{(1)}\,Q_{r}-\overline c_{r,s}^{(1)}\,\overline Q_{r}}{x^{s}},
\ee
where the expansion coefficients are now {\it antisymmetric}:
\be
c_{r,s}^{(1)} = 2\,\frac{1-(-1)^{r+s}}{2}\,\frac{s-r}{r+s-2}\  , \qquad \qquad
\overline c_{r,s}^{(1)} = -2\,\frac{1-(-1)^{r+s}}{2}\,\frac{r+s-2}{s-r}.  \la{ant}
\ee
These are the expressions for the phase coefficients that we announced in the Introduction,  see  eqs. \eq{crsintro} and \eq{crsbarintro}.
These coefficients  (only $c_{r,s}^{(1)}$  is actually  contributing) now lead precisely to the    string theory expression  
 $\delta E^{AdS_{3}}_1$ that we got in \rf{314}  for the  non-analytic term   in the  circular $SU(2)$ string case, i.e.
 (cf. \rf{acc}) 
\be
\delta E^{\rm AdS_{3}}_1 = \frac{1}{\pi\,\big[1+\frac{2\,Q_{2}}{(4\,\pi\,\J)^{2}})\big]}\mathop{\mathop{\sum_{r\ge 1}}_{s\ge r+1}}_{r+s\,\rm odd}
\Big(\frac{1}{4\,\pi\,\J}\Big)^{r+s}\,2\,\frac{s-r}{r+s-2}\,\Big(
Q_{r+1}\,Q_{s}-Q_{r}\,Q_{s+1}
\Big),
\ee
This suggests that  (\ref{eq:diff})  is responsible   for  the regularization mismatch term $\DE$ in \rf{de}. 
 A hint in this direction is  that the $G$ functions in  (\ref{eq:diff}) are evaluated at $y=1$
and this is the large $n$ region where regularization issues are relevant.


\

We can now  compute the scattering phases between magnons as in the final part of section 3.2 of \cite{Gromov:2007cd}.
To this aim, we identify the dressing phase contribution in (\ref{middle}) as 
\be
e^{i\,\widehat{\mathcal V}(x_{2,i})} = 
\prod_{k\neq j}^{K_{2}} \sigma^{2}(x_{2,j}, x_{2,k})\,\prod_{k}^{K_{\overline 2}}
\sigma^{-2}(x_{2,j}, x_{\overline 2,k})
=
\prod_{k\neq j}^{K_{2}} e^{i\,\vartheta(x_{2,j}, x_{2,k})}\,\prod_{k}^{K_{\overline 2}}
e^{-i\,\wt \vartheta(x_{2,j}, x_{\overline 2,k})}.
\ee
Using the discrete definition (\ref{eq:GH-defs}) of the function $G$ and integrating  over $y$
in  \rf{core}   we obtain
\ba
\vartheta(x, y) &=& -\frac{\wh \alpha(x)\,\wh \alpha(y)}{2\,\pi\,(x-y)^{2}}\Big[2\,\log\Big(\frac{x+1}{x-1}\,\frac{y-1}{y+1}\Big)
+2\,\frac{(x-y)\,(x^{2}+y^{2}-2)}{(x^{2}-1)(y^{2}-1)}\Big], \\
\wt \vartheta(x, y) &=& -\frac{\wh \alpha(x)\,\wh \alpha(y)}{2\,\pi\,(1-x\,y)^{2}}\Big[2\,\log\Big(\frac{x+1}{x-1}\,\frac{y-1}{y+1}\Big)
-2\,\frac{(x-y)\,(x^{2}y^{2}-1)}{(x^{2}-1)\,(y^{2}-1)}\Big] \ . \la{ttt}
\ea
We notice that\footnote{Remarkably, a similar relation was found   for the corresponding 
Pohlmeyer-reduced theories: a product of the two phase factors which appear in the  S-matrix of the 
reduced $AdS_3\times S^3$ theory gives the phase
factor  of the reduced $AdS_5\times S^5$ theory \cite{PRphase}. 
The presence of the two phase  factors  in the $AdS_3\times S^3$  case is    connected 
with the non-simple (product) structure of the supergroup  defining the corresponding   supercoset. 
We are grateful to B. Hoare for a related  discussion.}
\be
\vartheta(x,y)+\wt \vartheta(x,y) = \vartheta_{\rm AdS_5}(x,y),
\ee
where $\vartheta_{\rm AdS_5}$ is the  corresponding \adss    expression for the 1-loop phase 
(see  eq.(28)  in \cite{Gromov:2007cd})
\be
\vartheta_{\rm AdS_5}= -\frac{\wh \alpha(x)\wh \alpha(y)}{\pi}\,\Big[
\Big(\frac{1}{(x-y)^{2}}+\frac{1}{(xy-1)^{2}}\Big)\,\log\Big(\frac{x+1}{x-1}\frac{y-1}{y+1}\Big)
+\frac{2}{(x-y)(xy-1)}\Big].
\ee

\renewcommand{\theequation}{7.\arabic{equation}}
\setcounter{equation}{0}

\section{Non-analytic  term in the 1-loop energy of $SU(2)$ circular string \\
in $AdS_{3}\times S^{3}\times S^{3}\times S^{1}$}

In this section we  shall  compute the  1-loop energy correction  for the  $SU(2)$ circular string
 moving    in the  string model based on the $\alpha$-dependent  
 $AdS_{3}\times S^{3}\times S^{3}\times S^{1}$ background. 
We shall discuss the structure of the 
non-analytic contribution for  generic $\alpha$ as well as in the $\alpha\to 1$ limit
corresponding to \adt case.

\subsection{The classical solution}

We write the metric of $AdS_{3}\times S^{3}\times S^{3}\times S^{1}$ as
\be  &&
ds^{2} = ds_{AdS_{3}}^{2}+\frac{1}{\alpha}\,ds^{2}_{S^{3}_{+}}+\frac{1}{1-\alpha}\,ds^{2}_{S^{3}_{-}}
+d\psi^{2},\\
&&ds^{2}_{AdS_{3}} = d\rho^{2}-\cosh^{2}\rho\,dt^{2}+\sinh^{2}\rho\,d\phi^{2}, \ \ \ \ 
ds^{2}_{S^{3}_{\pm}} = d\gamma_{\pm}^{2}+\sin^{2}\gamma_{\pm}\,d\varphi_{1, \pm}^{2}
+\cos^{2}\gamma_{\pm}\,d\varphi_{2, \pm}^{2}.\no 
\ea
The embedding coordinates are 
\be
&&Y_{3}+i\,Y_{0}= \cosh\rho\,e^{i\,t}, \ \ \ \ \ \ \ \ 
Y_{1}+i\,Y_{2} =  \sinh\rho\,e^{i\,\phi}, \\
&&X_{1, \pm }+i\,X_{2, \pm } = \frac{1}{\sqrt\alpha}\,\sin\gamma_{\pm }\,e^{i\,\varphi_{1,\pm }}, \qquad
X_{3, \pm }+i\,X_{4, \pm } = \frac{1}{\sqrt\alpha}\,\cos\gamma_{\pm }\,e^{i\,\varphi_{2,\pm }},
\ee
We choose the following classical solution
\be
\rho=0, \ t=\kappa\,\tau, \quad \gamma=\frac{\pi}{4}, \ \ \ 
\varphi_{1, \pm}=w_{\pm}\,\tau+m_{\pm}\,\sigma, \ \ \ 
\varphi_{2, \pm}=w_{\pm}\,\tau-m_{\pm}\,\sigma,\qquad \psi=0.\la{76}
\ee
The Virasoro constraints give
\be
\kappa^{2} = \frac{w_{+}^{2}+m_{+}^{2}}{\alpha}+\frac{w_{-}^{2}+m_{-}^{2}}{1-\alpha}.\la{77} 
\ee
and the classical energy  is   $E_0 =\sql \kappa$. We  first  specialize to the following  case of
\be
w_{+}=\alpha\,\J, \qquad\qquad
w_{-}=(1-\alpha)\,\J 
 \ . \la{666}
\ee
The relation $w_{+}/w_{-} = \alpha/(1-\alpha)$ is  due to the requirement 
that in the point-like  $m_\pm =0$ limit  this    configuration  should reduce
  to the supersymmetric massless   BMN-like geodesic 
discussed in   \cite{Babichenko:2009dk}.
The solution then has two equal spins in each of the two spheres:
\beqa
	J_1 = J_2= \rl \,   \frac{w_+}{2\a} ={1 \ov 2}   \rl {\J}\ &\text{on }& S^3_+\  ,\ \\ \nn
	J_1 = J_2= \rl \,   \frac{w_-}{2(1-\a)} = {1 \ov 2}   \rl {\J} \ &\text{on }& S^3_- \ . 
\eeqa
Choosing further\foot{Below we  shall also consider  the  case with  non-zero $m_-$.}
\beq
\label{mmz}
	m_{+}=m,\qquad \qquad m_{-}=0\ ,
\eeq
we  get   in $S^3_-$ a single orbital momentum instead of two spins (by an $SO(4)$ rotation the solution on $S^3_-$ can be 
transformed into  geodesic   along   big circle). 
Since we  have two spins in $S^3_+$,  we  may   refer to this  case  as  an   $SU(2)$ solution.

\subsection{Fluctuation frequences}

Let us start with bosonic fluctuations. 
We find one massless and two massive fluctuations in $AdS_3$
\ba
\omega^{AdS_{3}\  (1)}_{n} =  n, \qquad 
\omega^{AdS_{3}\  (2,3)}_{n} =  \sqrt{n^{2}+\kappa^{2}}.
\ea
From the point of view of bosonic  fluctuations, the two 3-spheres are decoupled.  The characteristic equation for the  $S^3_\pm$ frequences is 
\be
\det\begin{pmatrix}
\omega^{2}-n^{2} & i\,(w\, \omega-m n) & i\,(-w\,\omega-m n) \\
-2\,i\,(w\, \omega-m n) & \omega^{2}-n^{2} & 0 \\
2\,i\,(w\, \omega+m n) & 0 & \omega^{2}-n^{2}
\end{pmatrix}
=0 
\ee
giving as in section 3  a massless and two massive modes 
\ba
\omega^{S^{3}_{\pm}\ (1)}_{n} = n, \qquad \qquad
\omega^{S^{3}_{\pm}\ (2,3)}_{n} = \sqrt{n^{2}+2\,w_\pm  ^{2}\pm 2\,\sqrt{n^{2}\,(w_\pm^{2}+m_\pm^{2})+w_\pm^{4}}}.
\ea
Finally, there  is also a massless mode from $S^1$
\be
\omega^{S^{1}}_{n} = n.
\ee
The discussion of  the fermionic fluctuations is similar, e.g., to the one in 
  \cite{Forini:2012bb}. The quadratic part of the GS Lagrangian reads
\be
L_{\rm GS} = i\,\Big(\sqrt{-h}\,h^{ab}\,\delta^{IJ}-\epsilon^{ab}\,\sigma_{3}^{IJ}\Big)\,
\overline\theta\,\rho_{a}\,\mathbf D_{b}^{JK}\,\theta^{K}, 
\ee
where $\rho_{a} = \partial_{a}X^{\mu}\,E^{A}_{\mu}\,\Gamma_{A}$, and
\be
\mathbf D_{b}^{JK}\,\theta^{K}=\delta^{JK}\,\Big(\partial_{b}+\frac{1}{4}\,\omega^{AB}_{\mu}\,\partial_{b}
X^{\mu}\,\Gamma_{AB}\Big)\,\theta^{K}+\frac{1}{24}\,F_{MNP}\,\Gamma^{MNP}\,\rho_{b}\,\sigma_{1}^{JK}
\,\theta^{K}.
\ee
The RR 3-form flux term here is 
\be
F_{MNP}\,\Gamma^{MNP} = 6\,\Big(\Gamma^{012}+\sqrt\alpha\,\Gamma^{345}+\sqrt{1-\alpha}\,\Gamma^{678}\Big) \equiv 6\,\overline\Gamma \ , 
\ee
where the 012, 345, 678, and 9 coordinates refer to the factors in $AdS_{3}\times S^{3}_{+}\times S^{3}_{-}\times S^{1}$, i.e. 
\be
\begin{array}{l|cccccccccc}
\mu & 0 & 1 & 2 & 3 & 4 & 5 & 6 &  7 & 8 & 9\\
X^{\mu} & t & \rho & \phi & \gamma_{+} & \varphi_{1, +}& \varphi_{2, +}
& \gamma_{-} & \varphi_{1, -}& \varphi_{2, -} & \psi 
\end{array}
\ee
Fixing $\kappa$-symmetry by $\theta^{1}=\theta^{2}=\theta$, we end with
\ba
L_{\rm GS} &=& -2\,i\,\overline\theta\,\Big(-\rho^{a}\,D_{a}-\frac{1}{4}\,\rho^{a}\,\overline\Gamma\,\rho_{a}\Big)\,\theta = -2\,i\,\overline\theta\,\mathbf{D}_{F}\,\theta, \\
D_{a} &=& \partial_{a}+\frac{1}{4}\,\omega^{AB}_{\mu}\,\partial_{a}
X^{\mu}\,\Gamma_{AB}.
\ea
The fermionic frequencies are the zeroes of the determinant of the fermionic operator. 
Let us define the two polynomials
\ba
&&P_{1}(\omega) =\frac{1}{16\,\alpha^{2}\,\kappa}\,\Big[\kappa  \Big(m^4+2 \alpha  m^2 \left((3-4 \alpha ) \mathcal{J}^2+(4-8 \alpha ) n^2\right)
 \no \\
&&  \ \ \ \ +\alpha ^2 \left((5-8 \alpha ) \mathcal{J}^4+16 n^4+8 \mathcal{J}^2 n^2\right)\Big) \no \\
&&\ \ \ \  -4 (\alpha -1) \mathcal{J}
   \left(\alpha  \mathcal{J}^2+m^2\right) \left(m^2+\alpha  \left(\mathcal{J}^2+4 n^2\right)\right)\Big]
   \cr 
   && \ \ \ \ +\omega  \Big((2 \alpha -1) \kappa  \mathcal{J}^2+\frac{(\alpha -1) \mathcal{J}
   \left(m^2+\alpha  \left(\mathcal{J}^2+4 n^2\right)\right)}{2 \alpha }\Big) \\
  &&  +\frac{\omega ^2 \left(m^2 (\kappa -2 \alpha  \mathcal{J}+\mathcal{J})-\alpha  (\mathcal{J}-\kappa ) \left(3 (2 \alpha -1)
   \mathcal{J}^2+4 n^2\right)\right)}{2 \alpha  (\mathcal{J}-\kappa )}+\omega ^3 (2 \mathcal{J}-2 \alpha  \mathcal{J})+\omega ^4,\no \\
 &&  P_{2}(\omega) = \frac{1}{16\,\alpha^{2}\,\kappa}\,\Big[
4 (\alpha -1) \mathcal{J} \left(\alpha  \mathcal{J}^2+m^2\right) \left(m^2+\alpha  \left(\mathcal{J}^2+4 n^2\right)\right)\nonumber \\
&&\ \ \ 
+\kappa  \left(m^4+2 \alpha  m^2 \left((3-4 \alpha ) \mathcal{J}^2+(4-8 \alpha )
   n^2\right)+\alpha ^2 \left((5-8 \alpha ) \mathcal{J}^4+16 n^4+8 \mathcal{J}^2 n^2\right)\right)\Big]
   \nonumber \\  &&\ \ 
   +\omega  \Big((2 \alpha -1) \kappa  \mathcal{J}^2-\frac{(\alpha -1) \mathcal{J}
   \left(m^2+\alpha  \left(\mathcal{J}^2+4 n^2\right)\right)}{2 \alpha }\Big) \\
   &&
   -\frac{\omega ^2 \left(m^2 (\kappa +(2 \alpha -1) \mathcal{J})+\alpha  (\kappa +\mathcal{J}) \left(3 (2 \alpha -1) \mathcal{J}^2+4
   n^2\right)\right)}{2 (\alpha  (\kappa +\mathcal{J}))}+\omega ^3 (2 \alpha  \mathcal{J}-2 \mathcal{J})+\omega ^4.\no 
 \ea
We can prove that the roots of $P_{1,2}(\pm \omega)=0$ are the distinct roots of 
\be
\det\Big(-\rho^{a}\,D_{a}-\frac{1}{4}\,\rho^{a}\,\overline\Gamma\,\rho_{a}\Big) = 0.
\ee
Choosing the signs of  $\omega_{n}^{F\,(1, 2, 3, 4, 5, 6, 7, 8)}$ such that for large $n$ we have $\omega_{n}^{F\,(i)} = n + \mc O(1)$, we can check that\footnote{The two zero modes come from one in $AdS_{3}$, one in $S^{3}_{+}$, 
one in $S^{3}_{-}$, one from $S^{1}$ minus two conformal-gauge  ghosts.
}
\be
\bar e(n) = \sum_{i=1}^{2}\omega_{n}^{AdS_{3}\,(i)}+\sum_{i=2}^{3}\omega_{n}^{S^{3}_{+}\,(i)}
+\sum_{i=2}^{3}\omega_{n}^{S^{3}_{-}\,(i)}+2\,|n|-\sum_{i=1}^{8}\omega_{n}^{F\,(i)} =
\mathcal{O}\Big(\frac{1}{n^{2}}\Big),
\ee
ensuring UV finiteness.

\subsection{Non-analytic part of $E_{1}$}

\def \dE  {\delta E_1}

The one-loop energy is 
\be
E_{1} = \sum_{n\in\mathbb Z} \, e(n), \qquad \qquad   e(n) = { \bar e(n) \ov 2 \kappa} \ , \ \ \ \ \ \ 
 \qquad \kappa = \sqrt{\J^{2}+\frac{m^{2}}{\alpha}}.
\ee
The non-analytic part can be found by as  discussed in \rf{310}
\be
\dE
 = \frac{\J}{2\,\kappa}\,\int_{-\infty}^{\infty}  dx \, \bar e(\J\,x) \ . 
\ee
From the  above  expressions  one  finds 
\be
&&\bar e(\J x) = \frac{m^{2}}{\alpha\,\J}\,\Big[\frac{\sqrt{2 \alpha ^2-2 \alpha  \sqrt{\alpha ^2+x^2}+x^2}-\sqrt{2 \alpha ^2+2 \alpha  \sqrt{\alpha ^2+x^2}+x^2}}{2 \sqrt{\alpha ^2+x^2}} \no \\ && \ \ \ \ \ \ \ \ \ \ \ \  +\ \frac{\alpha -1}{\sqrt{(\alpha -1)^2+x^2}} +\frac{1}{\sqrt{x^2+1}}\Big]+\dots.
\ee
Computing the integral, we find
\be
\dE
= \frac{m^{2}}{\alpha\,\J}\Big[\alpha\,\log \alpha +(1-\alpha)\,\log(1-\alpha)\Big]+\dots.
\ee
Going to the next order, and setting 
\be
\label{ldef}
L \equiv  \alpha\,\log  \alpha+(1-\alpha)\,\log(1-\alpha),
\ee
we find that for $0<\a<1$
\beq
\label{ecas}
	\dE=\frac{m^{2}\,L}{\alpha\,\J}
+\frac{m^{4}}{\alpha^{2}\,\J^{3}}\Big[\frac{1}{4}-\frac{L}{2}+\frac{1-\alpha}{8}\log\alpha+\frac{(\alpha-1)(\alpha+3)}{8\,\alpha}\log(1-\alpha)\Big] + ... \ , 
\eeq
while for $\a=1$
\beq
	\dE=\frac{m^{4}}{2\,\J^{3}}+\dots\ \ \ .
\eeq
%
The $\alpha=1$ case is in agreement with (\ref{314}). Actually, the limit $\alpha\to 1$ is discontinuous with a 
jump that is due to the extra massless modes that appear when $\alpha=1$.
Notice that the $\mathcal O(1/\J^{3})$ correction is not symmetric under $\alpha\to 1-\alpha$ since we have set $m_{-}=0$.

\subsection{Case of  $\alpha\to 1-\alpha$ symmetric solution
 and ``renormalization''  of string tension}

It  is  interesting to  consider the case  with manifest  symmetry under  $\alpha\to 1-\alpha$. To this end  we repeat
 the 1-loop calculation  assuming  that   instead of \rf{666}, \eq{mmz}  our classical solution now has\foot{We need then to   assume   that   $m_\pm$ are integers
 (which imposes a restriction on $\a$)   but 
  this is not important for the present computation.} 
\be w_+ = \a \J \ , \ \ \ \ \ \ \     w_-= (1-\a) \J\ , \qquad \qquad
m_{+}=\alpha\,m,\qquad \ \ \  m_{-}=(1-\alpha)\,m.
\ee
The calculation is completely 
similar, but the result is much simpler:
\be
\label{732}
\dE= \left\{
\begin{array}{ll}
\displaystyle
\frac{m^{2}\,L}{\J}
+\frac{m^{4}}{\J^{3}}\Big(\frac{1}{4}-\frac{L}{2}\Big)
+\frac{m^{6}}{24\,\J^{5}}\Big(-7+9\,L\Big)+
\dots, \ \ \ & 0<\alpha<1 \\ \\
\displaystyle \frac{m^{4}}{2\,\J^{3}}-\frac{7\,m^{6}}{12\,\J^{5}}+\dots,\ \   & \alpha=1
\end{array}\right. 
\ee

Let us  now show that the  $L=L(\alpha)$  dependent terms in  \rf{732}  can be removed by a coupling
 redefinition. Recall that  in the above expressions  we  used 
$\J = \frac{J}{\sqrt\lambda}$.   
Let   us  now introduce $h(\lambda)$ such that at strong coupling
\be
\label{hcirc}
h(\lambda) = \frac{\sqrt\lambda}{4\pi}+\c+\mc O\Big(\frac{1}{\sqrt\lambda}\Big),
\ee
and define  
\be \J_{h}=\frac{J}{4\,\pi\,h(\lambda)} \ . \ee 
The  classical plus one-loop energy corresponding to the case  of $m_{+}=\alpha\,m$ and  $m_{-}=(1-\alpha)\,m$
\be
E = \sqrt\lambda\,\sqrt{\J^{2}+m^{2}}+\Big[
\frac{m^{2}\,L}{\J}
+\frac{m^{4}}{\J^{3}}\Big(\frac{1}{4}-\frac{L}{2}\Big)
+\frac{m^{6}}{24\,\J^{5}}\Big(-7+9\,L\Big)+
\dots.
\Big]+\mc O\Big(\frac{1}{\sqrt\lambda}\Big)&&
\ee
can be expressed in terms of   $\J_{h}$ and  then expanded at large $h$. The choice
of\,\footnote{In \eq{hcirc} we defined $h(\lambda)$ so that  it has   $\frac{\rl}{4\pi}$ as the leading term  at strong coupling  by
 analogy with the \adss and \adst cases. Had we chosen it to be twice this value, i.e. 
  $h(\lambda)=\frac{\rl}{2\pi}+\c+O\Big(\frac{1}{\sqrt\lambda}\Big)$  as  in 
  \cite{OhlssonSax:2011ms}, we would get, instead of \eq{ahs1}, the relation $\c=\frac{L}{2\pi}$ which is consistent with
  what was found in \cite{Abbott} for the giant magnon case.}
\be
\label{ahs1}
\c_{_{AdS_3\times S^3 \times S^3\times S^1}} = \frac{L}{4\pi} 
\ee
removes all the  $L$-dependent (i.e. $\a$-dependent, with  $0 < \alpha < 1$)  terms  
in the 1-loop  energy
and we find 
\be
E = 4\,\pi\,h\,\sqrt{\J_{h}^{2}+m^{2}}+\Big(
\frac{m^{4}}{4\,\J_{h}^{3}}-\frac{7\,m^{6}}{24\,\J_{h}^{5}}+
\dots.
\Big)+\mc O\Big(\frac{1}{h}\Big).
\ee
Note that for the non-symmetric solution $\eq{mmz}$ this redifinition of the tension also removes the $L$-dependent part of the 1-loop energy \eq{ecas}  (although it  does not  eliminate  all of the dependence on $\a$).

It is natural to expect that this 
  effective string   tension  $h(\lambda)$ should be identified with the  interpolating coupling in the corresponding Bethe Ansatz.

\

\

\noindent 
{\bf Note added}:

\noindent 
While this paper was in preparation there appeared ref.\ci{Abbott} 
which also discussed 1-loop corrections to  some semiclassical string configurations in $AdS_3 \times S^3 \times S^3 \times S^1$.

\subsection*{Acknowledgments }

We would like to thank N. Gromov, B. Hoare, T. Klose, R. Roiban, B. Stefanski  and S. van Tongeren  for useful discussions. 
The work of F.L.-M. was supported in part by grants RFBR-12-02-00351-a, PICS-12-02-91052, by Russian Ministry of Science and Education under the grant 2012-1.1-12-000-1011-016 (contract number 8410), and by a grant of the Dynasty Foundation. F.L.-M. is also grateful for hospitality to Imperial College London where a part of this work was done. The work of AAT was supported by the STFC grant ST/J000353/1
and by the ERC Advanced grant No.290456.
 

\appendix
\numberwithin{equation}{section}

\setcounter{equation}{0}

\section{One-loop correction to energy of long folded  spinning string in \adt and \adst}
\label{sec:folded}

In this section we will study the  1-loop  correction to energy 
of long    
folded    spinning string   in the \adt and  \adst backgrounds.
 In both cases the solution has the same form as in \adss and describes a folded string carrying  large  AdS   spin 
  $S$ and and angular momentum  $J$  (for more details see 
  \cite{Frolov:2006qe,Forini:2012bb}). We 
 will consider the long string limit,
\beq
\label{longlim}
        S\gg J, \ \ \ \ \ \ \ \ \ \   \ x=\frac{\rl}{\pi J}\log{S}=\text{fixed},\ 
\eeq
in which the solution takes simple ``homogeneous'' form. 

Below we will show that for the \adt theory, matching with the one-loop energy of the folded string provides the same constraints on coefficients of the dressing phase in the ABA as the matching with the circular string discussed in section 3. We will also discuss the folded solution in the 
 \adst background, obtaining a closed expression for the one-loop energy which will allow us to analyze the structure of its non-analytic part.

\subsection{Bethe ansatz calculation in \adt  case}
The energy of the long folded string 
can be written in the form
\beq
\label{egenf}
	E=E_{0}+E_1=S+J\sqrt{1+x^2}+\frac{J}{\rl}F(x)+\dots\ , 
\eeq
where the first two terms are the classical energy $E_{0}$ and the third term is the one-loop correction $E_1$. This correction was computed for the \adt case in \cite{Forini:2012bb},
\beq
	E_1 =\frac{\J}{u} \Big[ \!\!
	  -\big(1-u^2\big) + \sqrt{1-u^2} - u^2\log u - (2-u^2)\log\Big(1+\sqrt{1-u^2}\Big)
	  \! \Big] \ ,
\eeq
where
\beq
	\J \equiv  \frac{J}{\rl}\ , \ \ \ \ \ \ \ \ \ \ \ \ u\equiv\frac{1}{\sqrt{1+x^2}} \ . 
\eeq
The ``non-analytic'' terms here, which in the \adss case were captured by the dressing phase in the ABA, are the terms with even powers of $x$ in the small $x$ expansion 
\beq
\label{foldedAdS3exp}
        \frac{1}{\J}E_1=-\frac{4 x^3}{3}+\frac{x^4}{2}+\frac{4 x^5}{5}-\frac{5
   x^6}{12}-\frac{64 x^7}{105}+\frac{17 x^8}{48}+\frac{32
   x^9}{63}-\frac{149 x^{10}}{480}+\dots\ . 
\eeq
Comparing with the \adss result \cite{Frolov:2006qe},
\beq
\label{ef5}
        \frac{1}{\J}E_1=-\frac{4 x^3}{3}+\frac{4 x^5}{5}+\frac{x^6}{3}-\frac{64
   x^7}{105}-\frac{2 x^8}{3}+\frac{32 x^9}{63}+\frac{43
   x^{10}}{40}+\dots\ , 
\eeq
we see that for \adt the non-analytic part starts one order earlier than in \adss case, 
 just as it happened for the circular string discussed  in section 3.

The one-loop energy of the long folded string in \adss was reproduced from the ABA in \cite{Casteill:2007ct}. The ABA equations for the $sl(2)$ sector of the \adt theory  are the same as in \adss up to the dressing phase, so the ABA prediction for the {\it analytic} part of the energy (sum of terms with even powers of $x$ 
in the small $x$ expansion) does not change, since it is only sensitive to the classical (AFS) part of the phase. In agreement with this prediction, the analytic part of the string result \eq{foldedAdS3exp} is the same as in the \adss case \eq{ef5}. For the {\it non-analytic} part the calculation of \cite{Casteill:2007ct} is straightforward to adapt to our case, and the dressing phase we propose leads to the following expression:
\beqa
	\frac{1}{\J}\delta E_1&=&\frac{1}{4} x^4 c_{1,2}^{(1)}+\frac{1}{16} x^6 (-4
  c_{1,2}^{(1)}+c_{1,4}^{(1)}-c_{2,3}^{(1)})\\ \nn
   & &\ + \ \frac{1}{64}
   x^8 \big(15 c_{1,2}^{(1)}-7 c_{1,4}^{(1)}+2 c_{1,6}^{(1)}+7 c_{2,3}^{(1)}
   -2 c_{2,5}^{(1)}+c_{3,4}^{(1)}\big)+\dots
\eeqa
Matching it with the even powers of $x$ in the string result \eq{foldedAdS3exp} we find exactly the same relations \eq{csu2f}-\eq{csu2l} for coefficients $c_{r,s}^{(1)}$ as for the circular string!

\subsection{One-loop correction to energy in \adst}

For the \adst background with generic $\a$ the fluctuation frequencies around the folded string solution in the limit \eq{longlim} were computed in \cite{Forini:2012bb}, and the 1-loop energy was obtained in closed form for special cases\footnote{Note that physical quantities, e.g. the one-loop energy, are symmetric under $\a\to1-\a$ as this is a symmetry of the background.} $\a=0$ and $\a=\frac{1}{2}$. Here we will compute it for generic $\a$, which will allow us to explore, in particular, the dependence on $\a$ of the non-analytic part of the energy.

The one-loop correction is defined by
\beq
\label{GenSumFolded}
	  E_1^\a = \frac{1}{2\kappa} \sum_{n=-\infty}^{\infty} \Big[
	  \sum_{i=1}^{8} \omega_i^B(n) - \sum_{i=1}^{8} \omega_i^F(n)
	  \Big]\,\ , 
\eeq
where in the regime we consider $\kappa=\frac{J}{\rl}\sqrt{1+x^2}\gg1$ and the frequencies are given in \cite{Forini:2012bb}
\begin{gather}
  \omega^B_{1,2}(n) = n \,, \qquad
  \omega^B_{3,4}(n) = \sqrt{n^2 + \alpha^2\J^2} \,, \qquad
  \omega^B_{5,6}(n) = \sqrt{n^2 + (1-\alpha)^2\J^2} \,, \\
  \omega^B_{7,8}(n) = \sqrt{n^2 + 2\kappa^2 \mp 2\sqrt{n^2\J^2 + \kappa^4}} \ ,
\end{gather}
\begin{equation}
  \omega^F_{1,2}(n) = \pm\frac{\J}{2} + n \,, \qquad
  \omega^F_{3,4}(n) = \pm\frac{\J}{2} + \sqrt{n^2 + \kappa^2} \ .
\end{equation}
The four other fermionic frequencies are given by the roots of two quartic equations\footnote{The corresponding 
equation in \cite{Forini:2012bb} contains a typo in the sign in front of $(\frac{1}{2}-\alpha)^2\J^2$ in the l.h.s.}
\beq
\label{fermEqFoldGena}
	  \Big[(\omega^F_i)^2 - n^2 - \big(\tfrac{1}{2}-\alpha\big)^2\J^2 \Big]^2 
	  = \kappa^2 \Big[
	  \omega^F_i +s \big(\tfrac{1}{2}-\alpha\big)\J\Big]^2 - (\kappa^2-\J^2) n^2  \ ,
\eeq
where $s=\pm1$. Note that the equation with $s=-1$ is obtained from the one with $s=+1$ by replacing $\omega\to-\omega$. At $\alpha=0$ the roots of this equation reduce to
\begin{equation}
  \omega^F_{5,6}(n) = \pm\frac{\J}{2} + n \,, \qquad\qquad 
  \omega^F_{7,8}(n) = \pm\frac{\J}{2} + \sqrt{n^2 + \kappa^2} \,
\end{equation}
in agreement with discussion of the spectrum in section C.4 of \cite{Forini:2012bb}.
 It is straightforward to check that the resulting  1-loop correction is UV finite.

To find   the 1-loop correction in the limit $\kappa\gg1$ one has to evaluate the integral
\beq
	\frac{1}{2\kappa}\int\limits_{-\infty}^{\infty}dn\Big[
	  \sum_{i=1}^{8} \omega_i^B(n) - \sum_{i=1}^{8} \omega_i^F(n)
	  \Big]\ , 
\eeq
which is nontrivial. The complication here is with the four fermionic frequencies that are solutions of the quartic equation -- they can be found as explicit but very involved functions of $n$. However,  let us make use of the fact that this quartic equation can be solved explicitly for $n(\omega)$ instead of $\omega(n)$. Then the trick is to use integration by parts: introducing a cutoff $\Lambda$ we get for these frequencies\footnote{Since $n$ enters the equation only as $n^2$ the frequencies are also even functions of $n$.} 
\beq
	\int\limits_{-\Lambda}^{\Lambda}dn\;\omega_i(n)=
	2\int\limits_{0}^{\Lambda}dn\;\omega_i(n)=
	2\(\omega_i n\)\Big|^{n=\Lambda}_{n=0}
	-2\int\limits_{\omega_i(0)}^{\omega_i(\Lambda)}d\omega_i\;n(\omega_i)\ . 
\eeq
After several changes of variables the integral over $\omega_i$ can be evaluated in elementary functions.\footnote{An important fact which turns out to reduce complexity of the integrand is that two of the roots $\omega$ of the quartic equation coincide when $n=0$.} The frequencies $\omega_i(0)$ can be found explicitly, and $\omega_i(\Lambda)$ are straightforward to find as an expansion at large $\Lambda$.

As a result, $E_1^\a$ is obtained in closed form:
\beqa
\label{eq:E1fa}
	\frac{1}{\J}E_{1}^{\alpha} &=& -\frac{ \sqrt{1-u^2} \sqrt{1-(1-2 \alpha )^2 u^2}}{2 u}+
	\frac{  \left(u^2-1\right)}{2 u}+\frac{  \sqrt{1-u^2}}{u}\nonumber \\
	&& +\frac{   \left(2 (1-\alpha ) \alpha  u^2-1\right)}{u}\,\log 2+
	\frac{  \left(u^2-2\right) \log \left(\sqrt{1-u^2}+1\right)}{u}\nonumber \\
	&& - \,u \Big[\alpha ^2 \log (\alpha )+(1-\alpha )^2 \log (1-\alpha )+
	(2 (\alpha -1) \alpha +1) \log u\Big] \nonumber \\
	&& -\frac{  \left(2 (\alpha -1) \alpha  u^2+1\right) \log \left(1-u^2\right)}{4 u}
	+ \sum_{i=1}^{2}\Big[f_{i}(\alpha)+f_{i}(1-\alpha)\Big],
\eeqa
where
\beqa
	f_{1}(\alpha) &=& \frac{(\alpha -1)   \left(\alpha  u^2-1\right) \log \left((1-2 \alpha ) u^2
	+\sqrt{\left(1-u^2\right) \left(1-(1-2 \alpha )^2 u^2\right)}+1\right)}{2 u}, \nonumber \\
	f_{2}(\alpha) &=& \frac{(\alpha -1)   \left(\alpha  u^2-1\right) \log 
	\left(\left(1-u^2\right) \sqrt{1-(1-2 \alpha )^2 u^2}+
	\sqrt{1-u^2} \left((1-2 \alpha ) u^2+1\right)\right)}{2 u}.\nonumber
\eeqa
This expression respects $\a\to1-\a$ symmetry and reduces at special values $\a=0$ and $\a=\frac{1}{2}$ to the 
expressions  found  in \cite{Forini:2012bb}.

An important outcome of this result  is the expression for the non-analytic part of the energy obtained from the small $x$ expansion of \eq{eq:E1fa}:
\beq
	\frac{1}{\J}\delta E^\alpha_1 =
	 Lx^2+\frac{1}{4} (-2 L+1)x^4
	+\frac{1}{24} (9L-5)x^6+\frac{1}{96} (-30L+17)x^8
	+\dots
\eeq
where $L$ is the same quantity that appeared for the circular string in \eq{ldef}. In complete analogy with the circular string case, all dependence on $\a$ is again removed by the same shift in the tension \rf{hcirc}. The reason for this is the relation
\beq
\label{eaa0}
	\delta E^\alpha_1=\frac{1}{2}\delta E^{\alpha=0}_1 +
	 L\J\,  x\frac{d}{dx}\sqrt{1+x^2} \ , 
\eeq
which also shows that, again, as $\alpha\to0$ the non-analytic part experiences a jump by a factor of two.
If we shift the tension in \eq{egenf} as
\beq
\label{shiftf}
	\rl\to\rl-4\pi \c
\eeq
while holding the charges $S$ and $J$ fixed, then the one-loop energy will get a contribution coming from the classical part $E_{0}$. The latter depends on the tension only through the variable $x$ (see \eq{egenf}), and we find that \eq{egenf} becomes
\beq
	E=S+J\sqrt{1+x^2}+\frac{J}{\rl}\(F(x)-4\pi \c\, x\frac{d}{dx}\sqrt{1+x^2}\)+\dots\ , 
\eeq
where the expression in the round brackets is  the modified one-loop energy. Then choosing
\beq
	\c=\frac{L}{4\pi}
\eeq
we see that due to \eq{eaa0} all terms with $L$ are removed. The shift \eq{shiftf} is equivalent to rewriting the string result in terms of the same interpolating coupling that was discussed in \eq{hcirc} for the circular string:
\beq
	h(\lambda)=\frac{\rl}{4\pi}+{ L\ov 4 \pi} +\O\big({ 1 \ov \sql} \big).
\eeq	
We can also compute the {\it analytic} part of the energy in the small $x$ expansion:
\beqa
	\frac{1}{\J}E_1^{\rm analytic}&=&\frac{4}{3} (R-1) x^3+\Big(-\frac{14 R}{15}+\frac{1}{30
   R}+\frac{4}{5}\Big) x^5
   \\ \nn
   & & +\ \Big(-\frac{1}{1120 R^3}
   +\frac{157
   R}{210}-\frac{13}{420 R}-\frac{64}{105}\Big)
   x^7+\dots\ , 
\eeqa
where
\beq
	R\equiv \sqrt{\a(1-\a)}\  .
\eeq
This result is valid for all $\a$ except $\a=0$ or $\a=1$ where it becomes singular. At these special values the analytic part can be found from \eq{foldedAdS3exp}.

\subsubsection{Large $\ell$ expansion}

Let us also discuss the expansion in terms of  $\ell$ defined by 
\be \ell=\frac{\pi \cJ}{\log S}=\frac{1}{x} \ee
  and the corresponding re-expansion at weak coupling (the corresponding expansion 
  for the \adss case is described in, e.g.,  \cite{Giombi:2010zi}).
The energy has the form
\be
	E=S +\frac{\rl}{\pi}{\rm f}(\ell,\rl)\log S+\dots\ , \ \ \ \ \ \ \ \ \ \ \ \ \ \ 
	{\rm f}(\ell,\rl)={\rm f}_0(\ell)+\ofrac{\rl}{\rm f}_1(\ell)+\dots
\ee
where  the classical part is the same as in \adss  case 
\beq
	{\rm f}_0(\ell)=\sqrt{1+\ell^2}\  ,
\eeq
while the 1-loop part can be found from \eq{eq:E1fa}.
With the aim of making a re-expansion at weak coupling (as in \cite{Giombi:2010zi}) we rewrite the energy as
\beq
	E=S+f(\lambda,\ell)\ln S+\dots\ , \ \ \ \ \ \ \ \ \ \ \ \
	f=\frac{{\rm f}(\ell,\rl)}{\ell}j\ , \ \ \ \ \ \   j=\frac{J}{\log S} = {\sql\ov \pi} \ell \ . 
\eeq
In the \adss case one obtains then   what looks like  a weak-coupling gauge theory  expansion:
\beqa
	f(\lambda,\ell)&=&
	\Big(j+\frac{\lambda}{2\pi^2j }-\frac{\lambda^2}{8\pi^4j^3}+
	\frac{\lambda^3}{16\pi^6j^5}+\dots\Big)
	\no \\
	& & + \ \Big(-\frac{4\lambda}{3\pi^3j^2}+\frac{4\lambda^2}{5\pi^5j^4}+
	\frac{\lambda^2\sqrt{\lambda}}{3\pi^6j^5}+\dots\Big)\ , 
\eeqa
where terms in the first line come from ${\rm f}_0$, and in the second line from ${\rm f}_1$. These two parts
 mix only at the order $\ofrac{j^5}$ where ${\rm f}_1$ provides a non-analytic contribution $\propto\lambda^{5/2}$, 
 so that  in the expansion
\beq
	f(\lambda,\ell)=j+\frac{c_{10}\lambda}{j}+\frac{c_{11}\lambda}{j^2}
	+\frac{c_{12}\lambda+c_{20}\lambda^2}{j^3}
	+\frac{c_{13}\lambda+c_{21}\lambda^2}{j^4}
	+\frac{p_5(\lambda)}{j^5}+\dots
\eeq
the coefficients $c_{ij}$ of lower-order terms should  be protected and independent of $\lambda$,
 while $p_5(\lambda)$    should be   a nontrivial interpolating function.

In \adst the tree-level part ${\rm f}_0$ is the same, while the 1-loop part  is different,  and  we find
\beqa
\label{fellads3}
	f(\lambda,\ell)&=&
	\Big(j +\frac{\lambda}{2\pi^2j }-\frac{\lambda^2}{8\pi^4j^3}+
	\frac{\lambda^3}{16\pi^6j^5}+\dots\Big)
	\\ \nn
	&+&\Big(
	\frac{\lambda^{1/2}  L}{\pi ^2 j}+
	\frac{4 \lambda  (R-1)}{3 \pi ^3
   j^2}-
   \frac{\lambda ^{3/2} (2 L-1)}{4 \pi ^4 j^3}+
   \frac{\lambda ^2
   \left(28 R+\frac{1}{R}+24\right)}{30 \pi ^5 j^4}+
   \frac{\lambda^{5/2}
(9 L-5)}{24 \pi ^6
   j^5}
	+\dots
	\Big)
\eeqa
The first non-analytic term $\frac{\lambda^{1/2}  L}{\pi ^2 j}$ here already appears at order $\ofrac{j}$ which is two orders earlier than in $AdS_5 \times S^5$.
 This suggests    that  in this case 
 there   should   be no protected coefficients at all, at least in the part with odd powers of $\ofrac{j}$.

This  conclusion, however,  changes if we  shift the tension as in \eq{shiftf} above. Then an extra contribution comes from
$\frac{\rl}{\pi}{\rm f}_0(\ell)\log S$, the terms with $L$ in \eq{fellads3} cancel and we get
\beqa
	f(\lambda,\ell)&=&
	\Big(j+\frac{\lambda}{2\pi^2j }-\frac{\lambda^2}{8\pi^4j^3}+
	\frac{\lambda^3}{16\pi^6j^5}+\dots\Big)
	\\ \nn
	&+&\Big(
	\frac{4 \lambda  (R-1)}{3 \pi ^3
   j^2}+
   \frac{\lambda ^{3/2} }{4 \pi ^4 j^3}+
   \frac{\lambda ^2
   \left(28 R+\frac{1}{R}+24\right)}{30 \pi ^5 j^4}-
   \frac{5\lambda^{5/2}
}{24 \pi ^6
   j^5}
	+\dots
	\Big)\ . 
\eeqa
Now   the first nontrivial term in $f(\lambda,\ell)$, i.e. the $\frac{\lambda}{j}$ term , appears to be protected.

\subsubsection{Subleading corrections in large $\kappa$}

Finally, we can also study subleading terms in the large $\kappa$ expansion of $E_1$, taking $J\to0$. In
the  \adss
we have
\beq
	E_1^{(0)}=\frac{1}{2\kappa}\sum\limits_{n=-\infty}^\infty
	\left[
	\sqrt{n^2+4\kappa^2}+2\sqrt{n^2+2\kappa^2}+5\sqrt{n^2}-8\sqrt{n^2+\kappa^2}
	\right] \ , 
\eeq
which gives  \cite{Frolov:2002av,SchaferNameki:2006gk, Beccaria:2010ry}
\beq
	E_1^{(0)}=-3\log2\; \kappa-\frac{5}{12\kappa}+\dots\ , 
\eeq
where  dots denote exponentially suppressed terms.
In the  \adst  case  setting $J=0$ (i.e. $\J=0$)  restricts solution to $AdS_3$,  the
fluctuations  frequencies become independent of  $\alpha$ and we find 
\beq
		E_1^{(0)}=\frac{1}{2\kappa}\sum\limits_{n=-\infty}^\infty
	\left[
	\sqrt{n^2 + 4 \kappa ^2}+3 \sqrt{n^2}-4 \sqrt{n^2 + \kappa ^2} 
	\right]\ . 
\eeq
Then using the Euler-Maclaurin formula we get
\beq
	E_1^{(0)}=-2\log2\; \kappa-\frac{3}{12\kappa}+\dots
\eeq
Terms of the type $\frac{k}{12\kappa}$ in expansion of $E_1^{(0)}$ come from the $k$ massless modes in the sum over $n$. In \adss the five bosonic massless modes give $\frac{5}{12\kappa}$, while in \adst the three massless modes (7 bosonic minus 4 fermionic) give $\frac{3}{12\kappa}$.


\end{document}